\documentclass{aa}  

\usepackage{graphicx}
\usepackage{txfonts}
\usepackage{bm}
\usepackage{xcolor}

\usepackage{longtable} 
\usepackage{pdflscape} 
\usepackage{multicol}
\usepackage{multirow}
\usepackage{booktabs}

\begin{document}

   \title{Oxygen-bearing organic molecules in comet 67P's dusty coma: first evidence for abundant heterocycles}
   \titlerunning{Oxygen-bearing organic molecules in comet 67P's dusty coma}

\author{N. H{\"a}nni\inst{1} \and K. Altwegg\inst{1} \and D. Baklouti\inst{2} \and M. Combi\inst{3} \and S. A. Fuselier\inst{4,5} \and J. De Keyser\inst{6} \and D. R. Müller\inst{1} \and M. Rubin\inst{1} \and S. F. Wampfler\inst{7}}

\institute{Physics Institute, Space Research \& Planetary Sciences, University of Bern, Sidlerstrasse 5, CH-3012 Bern, Switzerland.\\ \email{nora.haenni@unibe.ch}
\and
Institut d'Astrophysique Spatiale, Université Paris-Saclay, CNRS, Orsay, France.
\and
Department of Climate and Space Sciences and Engineering, University of Michigan, Ann Arbor, MI, USA.
\and
Space Science Division, Southwest Research Institute, San Antonio, TX, USA.
\and
Department of Physics and Astronomy, The University of Texas at San Antonio, San Antonio, TX, USA.
\and
Royal Belgian Institute for Space Aeronomy, BIRA-IASB, Brussels, Belgium.
\and
Center for Space and Habitability, University of Bern, Gesellschaftsstrasse 6, CH-3012 Bern, Switzerland.}

\date{Received XXX / Accepted YYY}


\abstract
{The puzzling complexity of terrestrial biomolecules is driving the search for complex organic molecules in the Interstellar Medium (ISM) and serves as a motivation for many in situ studies of reservoirs of extraterrestrial organics from meteorites and interplanetary dust particles (IDPs) to comets and asteroids. Comet 67P/Churyumov-Gerasimenko (67P) -- the best-studied comet to date -- has been visited and accompanied for two years by the European Space Agency's Rosetta spacecraft. Around 67P's perihelion and under dusty conditions, the high-resolution mass spectrometer on board provided a spectacular glimpse into this comet's chemical complexity. For this work, we analyzed in unprecedented detail the O-bearing organic volatiles. In a comparison of 67P's inventory to molecules detected in the ISM, in other comets, and in Soluble Organic Matter (SOM) extracted from the Murchison meteorite, we also highlight the (pre)biotic relevance of different chemical groups of species. We report first evidence for abundant extraterrestrial O-bearing heterocycles (with abundances relative to methanol often on the order of 10\% with a relative error margin of 30--50\%) and various representatives of other molecule classes such as carboxylic acids and esters, aldehydes, ketones, and alcohols. Like with the pure hydrocarbons, some hydrogenated forms seem to be dominant over their dehydrogenated counterparts. An interesting example is tetrahydrofuran (THF) as it might be a more promising candidate for searches in the ISM than the long-sought furan itself. Our findings not only support and guide future efforts to investigate the origins of chemical complexity in space, but also they strongly encourage studies of, e.g., the ratios of unbranched vs. branched and hydrogenated vs. dehydrogenated species in astrophysical ice analogs in the laboratory as well as by modeling.}


\maketitle

\section{Introduction}\label{sec:intro}
Along with meteorites \citep[e.g.,][]{jenniskens2000} and Interplanetary Dust Particles \citep[IDPs; e.g.,][]{levasseur-regourd2018}, comets are considered to be the major source of pristine organic matter delivered to the early Earth and hence may have played an important role in prebiotic chemistry and the processes that led to carbon-based life on Earth \citep[e.g.,][]{chyba1992}. Using observations at comet 67P/Churyumov-Gerasimenko (hereafter 67P), \citet{rubin2019a} have estimated that comets, when bringing 22\% of the xenon to Earth as proposed by \citet{marty2017}, also bring about the terrestrial biomass equivalent (or more) in the form of organic molecules. Assuming full conservation of the cometary organics budget during impact, this corresponds to about 17\,000 to 350\,000 impacts of 67P-like objects. For the main building blocks of life, which are lipids, amino acids, nucleic acids, and sugars, see, e.g., \citet{kwok2016}, so-called heteroatoms (i.e., atoms other than C and H, which can be O, N, S, or P) are of outstanding importance. Because of its abundance in space and in biomolecules like amino acids, fatty acids, and sugars, the heteroelement O has received special attention in astrochemistry. Extensive efforts have been made towards the detection and study of O-bearing molecules with possible roles in prebiotic chemistry.\\
The list of O-bearing molecules detected in the interstellar medium (ISM) is relatively short. As of mid-2021, \citet{mcguire2022} reports 22 neutral O-bearing organic molecules (radicals and disputed detections excluded) that are composed of 4 or more atoms. The most complex molecules among them are the structural isomers ethyl formate, methyl acetate, and hydroxy acetone with a molecular weight of 74 Da and a chemical sum formula equal to C$_3$H$_6$O$_2$. In contrast, a plethora of O-bearing molecules and macromolecules are known to be present in meteorites and, as recently published, also in samples returned from the near-Earth carbonaceous asteroid (162173) Ryugu \citep{naraoka2023,yabuta2023}. Oxygen is the most abundant heteroelement in both soluble organic matter \citep[SOM; e.g.,][]{botta2002,schmitt-kopplin2010} as well as insoluble organic matter \citep[IOM; e.g.,][]{alexander2017}. Both in IOM and SOM, the heteroelement oxygen is roughly an order of magnitude more abundant than the heteroelement nitrogen. Ryugu SOM, where only two carboxylic acids (formic and acetic acid) were detectable apart from various amines, amino acids, and N-containing heterocycles, seems to be an exception \citep{naraoka2023}. \citet{naraoka2023} explain the lack of O-bearing volatile molecules with the fact that the N-bearing species they find (with basic N functionality) may have been bound in the form of ammonium salts as already reported for Ceres \citep{desanctis2016} and comet 67P \citep{altwegg2020,poch2020,altwegg2022}. However, as we will discuss in more detail below, the extraction protocol may significantly influence quantitative results. Even less abundant than nitrogen in IOM and SOM is sulfur or phosphorus. A similar heteroelement abundance distribution relative to carbon has recently been reported for the volatile organics with molecular weights of up to 140 Da in comet 67P's dusty coma \citep{haenni2022}. Those complex organics expose an average sum formula of C$_1$H$_{1.56}$O$_{0.134}$N$_{0.046}$S$_{0.017}$. Notably, no organic source of phosphorus has been identified in comet 67P to date.\\
We use the same high-resolution mass spectrometric data set as in \citet{haenni2022}, which was collected on 3 August 2015 at comet 67P by the Rosetta Orbiter Spectrometer for Ion and Neutral Analysis (ROSINA) instrument suite \citep{balsiger2007}, and present an extensive study of this comet's inventory of volatile O-bearing organic molecules, which is still poorly constrained. Previous work was focusing on data from the time when the comet was far from its perihelion and activity was low and, hence, only a limited number of species -- the rather small and abundant ones -- were detected in 67P's coma \citep{leroy2015,altwegg2017,schuhmann2019a,rubin2019b}. The conditions on 3 August 2015, shortly before perihelion, when the data set analyzed for this work was collected, were distinctly different. Comet 67P was reaching its perihelion at 1.24 au from the Sun in early August 2015 and the coma was dusty. Consequently, enhanced desorption of heavier and thus more complex molecules can occur from ejected dust particles. Decoupled from the heat sink of the cometary surface, the temperatures of ejected particles can rise up to a few hundreds of Kelvin \citep{lien1990} in sunlight, whereas comet surface temperatures around that time were observed to not exceed 240 K \citep{tosi2019}.
For the most abundant subgroup of signals of pure hydrocarbon species in the 3 August 2015 data set, \citet{haenni2022} demonstrated a way to decompose the observed signals into contributions of individual neutral molecules. Following the same methodological approach, we present here an attempt to disentangle the second-most abundant group of cometary volatile organics, which is that of the O-bearing C$_n$H$_m$O$_x$ hydrocarbons, where n = 1--8, m = 0--14, and x = 1--2. After giving the relevant information on data acquisition as well as data reduction and interpretation, we compare and contrast our findings with other comets, meteoritic organic matter, and the ISM.

\section{Instrumentation and method}\label{sec:instr}

\subsection{Data acquisition}\label{subsec:data-ac}
ROSINA's high-resolution Double Focusing Mass Spectrometer (DFMS; \citet{balsiger2007}) was designed to detect the cometary volatiles originating from both the cometary bulk material and the ejected particles. In the neutral gas mode of the instrument, neutral species were allowed to enter the ionization chamber where they were ionized by electrons impacting with energies of 45 eV. The electron impact ionization (EII) process usually yields singly charged cations of the parent molecules also known as molecular ions (M). However, doubly charged species can form and the parent species can fragment during the EII process into fragment species. Note that for better readability and to avoid confusion with ions naturally present in the cometary coma, we avoid indication of the positive charges induced by the EII process. All molecular ions and charge-retaining fragments were subsequently extracted from DFMS' ionization chamber and transferred via an electrostatic analyzer and a sector magnet -- where they were separated according to their mass-to-charge ratios (\textit{m/z}) -- onto a stack of two Micro Channel Plates (MCPs). The MCPs were mounted in a Chevron configuration and released an electron cascade upon impact of analyte ions. The charge was then collected on two redundant rows of position sensitive Linear Electron Detector Array (LEDA) anodes with 512 pixels each. To scan the whole mass range analyzed in this paper, the DFMS had to measure the \textit{m/z}-range around each integer mass number separately and sequentially, which, including overhead time, took roughly 30 s per single mass spectrum. The maximally accessible \textit{m/z}-range was 12--180 in the late mission phase, while on 3 August 2015 only \textit{m/z} = 13--140 was scanned using the following combination of measurement modes: 222, 562, and 564. Those modes allowed scanning the \textit{m/z}-ranges \textit{m/z} = 44--80 between 15:03--15:20 UTC, \textit{m/z} = 80--140 between 15:33--16:00 UTC, and \textit{m/z} = 13--43 between 16:25--16:39 UTC. For \textit{m/z} > 70, a post-acceleration potential of 1000 V was applied in front of the detector to increase the detection efficiency for heavy species.\\
Although observed at other instances \citep{altwegg2017,altwegg2020}, on 3 August 2015 there was no sign of particles sublimating inside or close to the DFMS' ionization chamber, such as a sudden increase of the signal strengths or a blockage of electron or ion currents. It is important to note, however, that even these dust events do not result in relevant energetic processing of the analyte prior to EII as the particle's relative velocity is merely a few meters per second. The detected species are thought to be present in their free form either in the ice- or dust-dominated matrix from which they thermally desorb. \citet{haenni2022}, furthermore, argued in detail that the observed species are likely to have origins in the early history of our Solar System and are not decomposition products of cometary macromolecular matter \citep{fray2016}.

\subsection{Data reduction}\label{subsec:data-red}
\begin{figure}
   \centering
   \includegraphics[width=8cm]{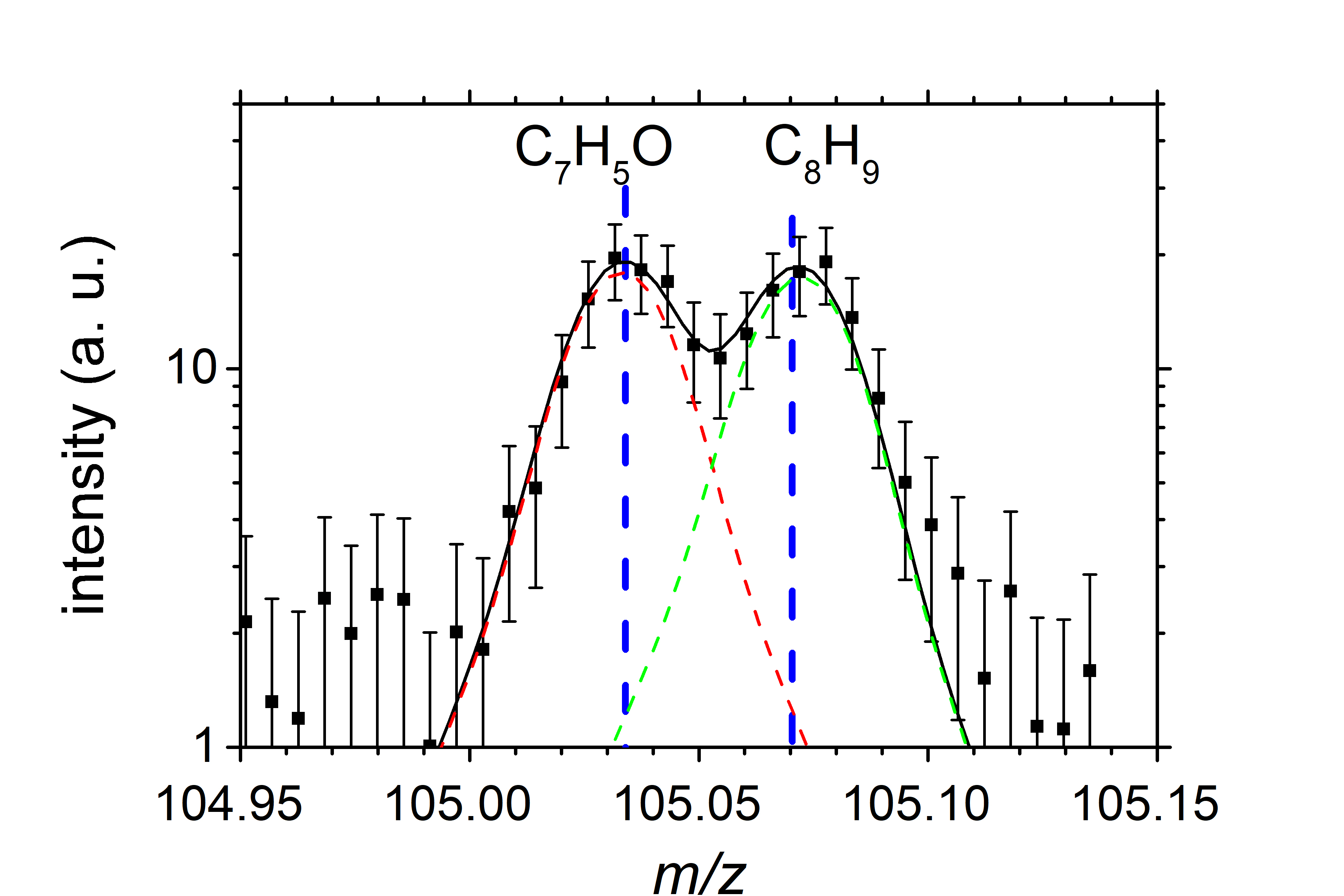}
      \caption{DFMS mass spectrum collected on 3 August 2015 around integer mass 105 (black square markers). The peaks -- here associated to C$_7$H$_5$O and C$_8$H$_9$ at the exact positions indicated by the dashed blue lines -- are fitted with double-Gaussian functions (dashed red and green lines). The standard deviation of the fitted peak position is $\approx 6\times 10^{-4}$.}
         \label{fig:spec}
   \end{figure}

The DFMS achieved a mass resolution of m/$\Delta$m = 3000 considering the full width at 1\% of the peak height of the molecular nitrogen peak at \textit{m/z} = 28 \citep{balsiger2007}. This resolution is sufficient to distinguish pure hydrocarbons from heteroatom-bearing hydrocarbons, which we exploit in the present analysis. As an example, Fig. \ref{fig:spec} shows the mass spectrum collected around \textit{m/z} = 105. The two registered signals, associated with C$_7$H$_5$O and C$_8$H$_9$, are well separated. See also supplementary Figure 2 in \citet{haenni2022}, which shows the mass spectrum around \textit{m/z} = 109 with three well-separated signals of C$_5$H$_3$NS, C$_7$H$_9$O, and C$_8$H$_{13}$. The relative peak position is defined once two or more signals are present in the mass spectrum. The absolute peak position and, hence, the mass scale, is defined by one correct assignment of a species (i.e., its exact mass) to a peak in the mass spectrum, usually of C$_n$H$_m$ and C$_n$H$_m$X species. Continuation of the homologous series in the adjacent mass spectra as C$_n$H$_{m\pm1}$ and C$_n$H$_{m\pm1}$X is frequently observed and helpful for verification of the mass scale. In addition, regularly measured major cometary species like H$_2$O, CO, CO$_2$, OCS, or CS$_2$ are used to define the mass scale. Selected candidate molecules are also verified by relative elemental and isotopic abundance considerations \citep{haenni2022,rubin2019b,lodders2021}. The signal intensities are extracted from fitting a double-Gaussian peak profile to each mass peak, see, e.g., \citet{leroy2015}, and corrected for the mass-dependent instrument sensitivity. However, for low-intensity signals with amplitudes of less than an order of magnitude above background level -- such as those of some heavier heteroatom-bearing hydrocarbon species--, the second Gaussian disappears in the background. After fitting the spectra and summing both LEDA rows, the obtained intensities were corrected for a drift in the signal strength. As detailed in \citet{haenni2022}, this drift originated from the combination of spacecraft slew, comet rotation, variability in the coma composition, as well as thermal changes that occurred during the time period of interest. The drift was corrected using repetitively measured mass ranges. The estimated errors on the extracted intensities of C$_n$H$_m$O signals are $\approx$30\% and those of C$_n$H$_m$O$_2$ signals are $\approx$50\%. This grouping is assumed for simplicity. However, there are slight differences for each individual species. These overall errors include:

   \begin{enumerate}
	    \item Statistical error: For C$_n$H$_m$O species, which are more abundant, the statistical error is a minor contribution and can be neglected while for C$_n$H$_m$O$_2$ species, which are less abundant, the statistical uncertainty is roughly $\approx$20\%.
	    \item Error related to detector gain: All species were measured with the highest gain step which means that for relative abundances related uncertainties cancel out.
	    \item Fit error and individual pixel gain error: Together, they are estimated to be $\approx$15\%.
	    \item Residual systematic error related to the drift: This error is estimated to be $\approx$15\%.
   \end{enumerate}
				
\noindent To derive the absolute abundance of contributing molecules, their fragmentation patterns must be taken into account. Here, this was done based on the reference mass spectra from NIST, which are obtained for 70 eV electron impact. Therefore, the measured signals have to be corrected for the ionization cross section of the analyte molecules as well as a species-dependent sensitivity, accounting for the variable sizes, masses, and forms of the ions. Unfortunately, data for the cross sections are available in literature only for a small set of simple molecules. The same accounts for the species-dependent sensitivities, which have been studied for the DFMS only for a limited set of the molecules investigated in this work. As these factors are not taken into account, individual abundances are given in arbitrary units. For the purpose of inter-comparison, it is most suitable to consider not the absolute, but rather the relative abundances of a specific species. However, normalization with respect to water, the major cometary volatile, is not suitable for this investigation. Data used here were collected during a time period where sublimation from ejected dust grains was strongly enhanced in comparison to sublimation from the cometary bulk material and water is expected to be depleted. The complex organic species seem to be correlated rather with methanol, cf. \citet{rubin2023} or Figure 5 in \citet{laeuter2020}. Consequently, we normalized our abundance values to methanol, which, itself, does have a negligibly small error. The methanol signal stands alone and is fitted well, leading to a statistical error $<$1\%. The individual pixel gain error is $\approx$5\%. Combining the errors of the other species as discussed above, the errors on the abundances relative to methanol are $\approx$30\% for C$_n$H$_m$O molecules and $\approx$50\% for C$_n$H$_m$O$_2$ molecules.

\subsection{Deconvolution of fragmentation patterns}\label{subsec:data-decon}
The so-called EII fragmentation pattern of a given molecule is the characteristic fingerprint of signals registered when the \textit{m/z}-range is scanned while the parent and fragment ions are being collected. It is usually graphed in terms of intensity on the y-axis and \textit{m/z} on the x-axis. Such mass spectra are collected in reference data bases like that from the National Institute of Standards and Technology \citep[NIST;][]{NIST}, which is the most extensive one as far as we know. The electron energy used to acquire NIST standard mass spectra is usually 70 or 75 eV which is different from the 45 eV used by the DFMS. However, the deviations in the fragmentation patterns resulting from this difference are minor \citep{schuhmann2019a,schuhmann2019b} and the related error is small compared to other sources of uncertainty. Furthermore, the NIST reference spectra only come in unit-resolution. This results in ambiguities for certain signals as to whether the O-atom is retained in the fragment species (O-bearing fragment) or not (pure hydrocarbon fragment) because both fragment types could and in some cases do exist. In the following, we briefly describe, for the most important O-bearing chemical functional groups, how the low-resolution mass spectra available from NIST are used to interpret our high-resolution data (here and in the following we use '--' to indicate chemical bonds and '-' to indicate that an atom/group of atoms was lost from the molecular ion):

   \begin{itemize}
	    \item Carboxylic acids (R--COOH, where R is a residual): Usually, carboxylic acids produce a detectable molecular ion signal (M signal) and they commonly fragment under loss of the hydroxy group, leading to a prominent M-OH signal.
	    \item Carboxylic acid esters (R--COO--R): Esters usually fragment due to bond cleavage next to the carboxyl function, i.e., the alkoxy group (--OR) is lost while hydrogen atoms are rearranged. M is mostly visible but weak.
	    \item Aldehydes (R--CHO): For aldehydes, M is observable while M-H is mostly weak or even absent. Fragmentation typically occurs under retention of the O atom while rather a part of R is lost, e.g., a methyl group leading to an M-CH$_3$ signal.
	    \item Ketones (R--CO--R): For ketones, major fragments originate from cleavage of the CC bond cleavage adjacent to the carboxyl function. The M signal is usually observed but rather moderate. 
	    \item Alcohols (R--OH): When alcohols fragment, favorably, water is lost, leading to a pure hydrocarbon signal at M-H$_2$O. Especially for primary alcohols, the resonance-stabilized CH$_3$O cation on \textit{m/z} = 31 is energetically favorable and usually of high intensity.
	    \item Ethers (R--O--R): For ethers, usually the main species are M and M-H.
   \end{itemize}
				
\noindent The loss of a water molecule from the molecular ion, leading to an M-H$_2$O signal, is common not only for molecules containing a hydroxyl function, such as alcohols or carboxylic acids, but it is also observed for aldehydes as well as cyclic and straight chain aliphatic ketones (see \citet{yeo1969} and literature cited therein). The rules above may not perfectly capture the fragmentation behavior of every molecule with the respective functional group, but they may well serve as rules of thumb and help to decide cases, where relevant signals are ambiguous in low-resolution NIST reference spectra.\\
When the analyte is a complex mixture of species, like for instance the cometary coma, then the characteristic fragmentation patterns of all species present in that mixture overlap, i.e., multiple species can contribute to a given mass peak. As a result, a sum fragmentation pattern is observed. To determine which species contribute, the sum fragmentation pattern must be deconvolved. \citet{haenni2022} demonstrated for the dominant subgroup of pure hydrocarbon species, C$_n$H$_m$, how the empirical principle of Occam's Razor \citep[e.g.,][]{clauberg1654} can be employed to find a solution to this problem. Here, we apply the same method to the subgroup of C$_n$H$_m$O$_x$ species (parents and fragments), which depicts the second-most abundant group of species, roughly half as abundant as the pure hydrocarbons.\\
Such a deconvolution attempt comes with a number of caveats: First, the fragmentation pattern has been experimentally determined only for a fraction of the large number of complex organic molecules in our mass range of interest. Second, even if reference spectra are available, some molecules do not yield significant M signals and thus cannot be unambiguously identified. This is especially the case for longer carbon chain species like for example pentanoic acid, pentanal, or 1-pentanol. Third, structural isomers possess the same exact mass and sometimes have very similar fragmentation patterns, which could render them indistinguishable by mass spectrometry. It is clear that larger molecules have more structural isomers (including exotic ones) and not for all of those isomers the mass spectra have been measured. Consequently, the selection of candidates becomes more ambiguous for larger molecules. It could be assumed, however, that the isomers with experimental data tend to be more stable (notably under terrestrial conditions) and, hence, might be more likely present in comets too.\\
Moreover, we have based our selection on arguments of plausibility. If a complex organic molecule has a very small M reference signal, the amount needed to explain the intensity observed by the DFMS may be very large, i.e., of the order of magnitude of very abundant cometary species like CO and CO$_2$. Even if such a molecule could not be discarded based on overshooting fragment signals, we still excluded the species based on the improbably high abundance. Even though additional molecules are likely present in the cometary coma beyond those captured by our deconvolution attempt, their total amount is limited by the uncertainties in the measured data. There are also cases where two structural isomers are very likely present in the cometary coma but their abundance ratio is badly constrained based on our data (e.g., glycolaldehyde and methyl formate, see below under \ref{sec:res}).
To sum up, the solution we present in the following is the most likely solution obtained from applying Occam's razor to the available reference data while taking into consideration also knowledge on cometary molecules. However, we cannot exclude a scenario where many more molecules (i.e., structural isomers of the candidates we selected) contribute to the observed overall fragmentation pattern. They could either add to the overall fragmentation pattern within the estimated error margins or (partially) replace some of the molecules, leading to decreased individual abundances of some of the larger candidate molecules. Consequently, our Occam's razor-based method might underestimate the diversity of O-bearing cometary species especially with higher molecular weights. We include these considerations when discussing our interpretation of the measured DFMS mass spectra below.


\section{Results}\label{sec:res}
\begin{figure*}
   \centering
   \includegraphics[width=\textwidth]{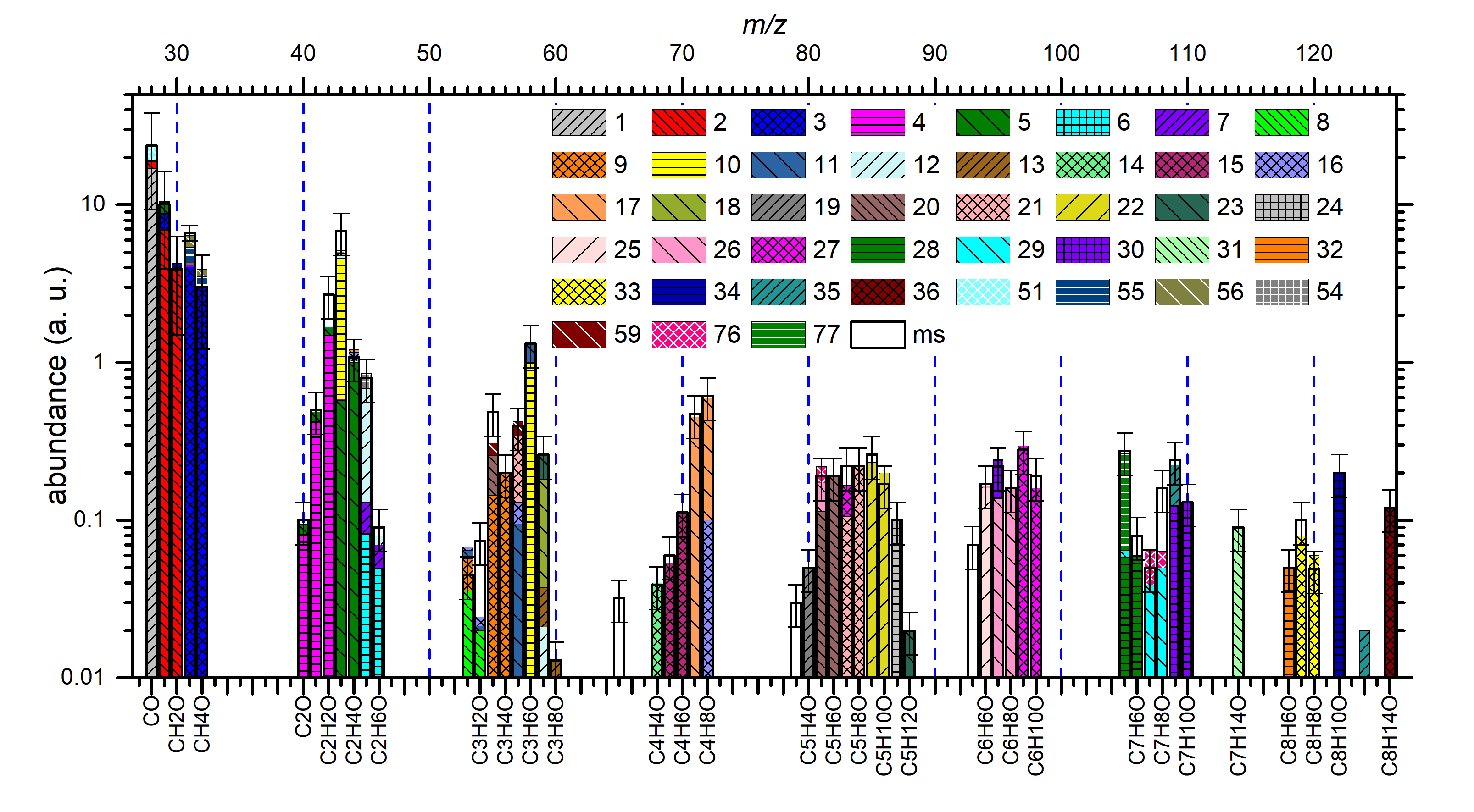}
      \caption{Occam's razor-based deconvolution of the subset of signals associated to C$_n$H$_m$O species as detected by the DFMS on 3 August 2015. The measured signals (ms) are given in arbitrary units (a. u.) with 30\% error margins. Color-coded we show the different contributions of the individual molecules to the observed sum fragmentation pattern of C$_n$H$_m$O species, see also details in the main text. The following 36 molecules have been selected: (1) carbon monoxide, (2) formaldehyde, (3) methanol, (4) ketene, (5) acetaldehyde, (6) dimethyl ether, (7) ethanol, (8) 2-propynal, (9) 2-propenal, (10) acetone, (11) propanal, (12) isopropanol, (13) \textit{n}-propanol, (14) furan, (15) 2,3-dihydrofuran, (16) butanal, (17) tetrahydrofuran, (18) 2-butanol, (19) 2,4-cyclopentadiene-1-one, (20) 2-methylfuran, (21) 2,3-dihydro-4-methylfuran, (22) tetrahydropyran, (23) 1-ethoxypropane, (24) 2-methoxy-2-methylpropane, (25) phenol, (26) 2,5-dimethylfuran, (27) 3-methoxycyclopentene, (28) benzaldehyde, (29) benzyl alcohol, (30) 2,3,5-trimethylfuran, (31) 2-methylcyclohexanol, (32) benzofuran, (33) 4-methylbenzaldehyde, (34) ethoxybenzene, (35) 6-methyl-3,5-heptadien-2-one, (36) 2,6-dimethylcyclohexanone, (51) carbon dioxide, (54) acetic acid, (55) glycolaldehyde, (56) methyl formate, (59) propanoic acid, (76) hydroquinone, (77) benzoic acid. For referencing and to distinguish the two populations, selected C$_n$H$_m$O molecules have been given numbers below 50 while C$_n$H$_m$O$_2$ molecules were given numbers above 50. Some of the latter produce relevant C$_n$H$_m$O fragments which show here and are thus listed.}
         \label{fig:CHO}
   \end{figure*}

\begin{figure*}
   \centering
   \includegraphics[width=\textwidth]{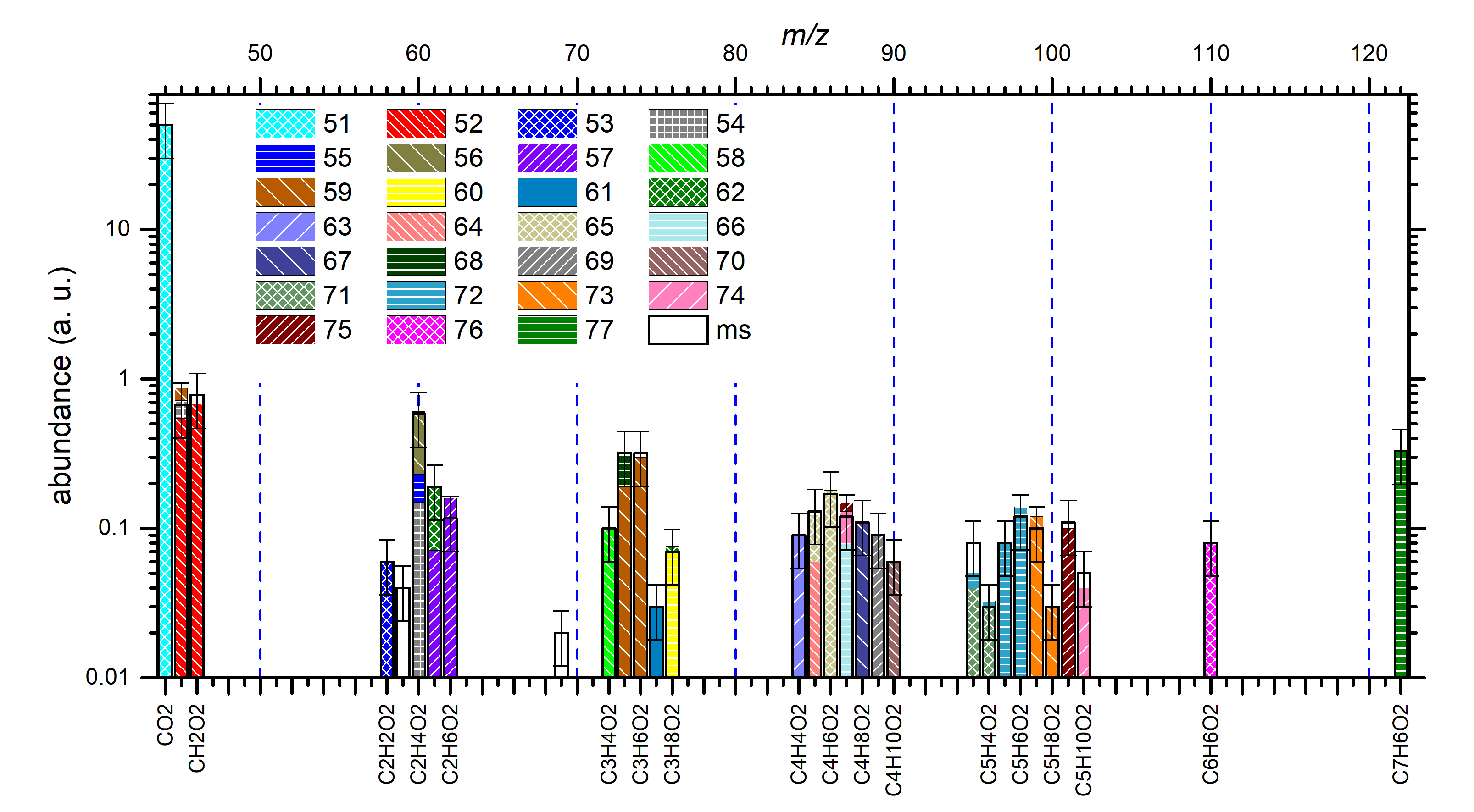}
      \caption{Occam's razor-based deconvolution of the subset of signals associated to C$_n$H$_m$O$_2$ species as registered by the DFMS on 3 August 2015. The measured signals (ms) are given in arbitrary units (a. u.) with 50\% error margins. Color-coded we show the different contributions of the individual molecules to the observed sum fragmentation pattern of C$_n$H$_m$O$_2$ species, see also details in the main text. The following 27 molecules have been selected: (51) carbon dioxide, (52) formic acid, (53) glyoxal, (54) acetic acid, (55) glycolaldehyde, (56) methyl formate, (57) ethylene glycol, (58) 2-propenoic acid, (59) propanoic acid, (60) 2-methoxyethanol, (61) methylal (62) 1,2-propanediol, (63) 2(3H)-furanone, (64) cyclopropanecarboxylic acid, (65) $\gamma$-butyrolactone, (66) 1,3-dioxane, (67) 1,4-dioxane, (68) butanoic acid, (69) 1-ethoxy-1-methoxyethane, (70) diethyl peroxide, (71) 3-furaldehyde, (72) 2-furanmethanol, (73) cyclopropanecarboxylic acid methyl ester, (74) propanoic acid ethyl ester, (75) 4-methyl-1,3-dioxane, (76) hydroquinone, (77) benzoic acid.}
         \label{fig:CHO2}
   \end{figure*}

Our Occam's razor-based deconvolution of the C$_n$H$_m$O$_{1-2}$ species observed on 3 August 2015 in comet 67P's dusty coma is shown in Figs. \ref{fig:CHO} and \ref{fig:CHO2} and is explained subsequently. Table \ref{tab:overview} lists all selected candidate molecules, grouped according to their primary chemical functionality. CO and CO$_2$ are listed as reference species. To present our results in an understandable and compact way to the reader, structural isomers are combined and listed according to increasing molar masses. Moreover, we indicate which isomers have been observed in comets \citep{biver2019} and in the ISM \citep{mcguire2022} because we sometimes consider those identifications to support the interpretation of our data. Note that for simplicity we do not reference the original literature which, however, are found in the respective review papers. For each of the selected candidate molecules, we denote the characteristic signals used for the identification and list the estimated abundance relative to methanol (ARM) in percent. We discuss possible alternative isomers and assign every identification with a level-of-confidence indicator (LCI): 1 = maybe present, 2 = likely present, 3 = with great certainty present. However, our LCIs are based on the available reference data and knowledge on cometary and ISM species. For example, if a specific molecule is the only isomer with reference data in NIST and is in suitable agreement with our data it will get an LCI of 3, despite that it is possible that other isomers without mass spectrum in NIST would even produce a better match.\\
After this procedure, some observed intensity remains unexplained, i.e., blank in Figs. \ref{fig:CHO} and \ref{fig:CHO2}. This intensity is mostly associated with species with a low number of hydrogen atoms (compared to the fully saturated molecule): C$_3$H$_2$O, C$_4$HO, C$_5$H$_3$O, C$_6$H$_5$O, C$_2$H$_3$O$_2$, and C$_3$HO$_2$. Some of those species might be radicals for which no NIST reference data exists. \citet{haenni2022} observed a similar situation for the pure hydrocarbon species and provided additional comments. Once the full cometary inventory of complex organics is done, these residual intensities can be investigated systematically.

\subsection{C$_n$H$_m$O species}\label{subsec:CHO}

   \begin{itemize}
	    \item CO (\textit{m/z} = 28): Carbon monoxide (no. 1; LCI = 3; ARM = 292), here identified via M, was reported to be present in comet 67P previously \citep{leroy2015,rubin2019b}, studied in situ in great spatial and temporal detail with multiple instruments \citep{biver2019b,laeuter2020,combi2020}, and observed remotely in many comets \citep{ahearn2012,dellorusso2016} as well as in the ISM \citep{mcguire2022}.
	    \item CH$_2$O (\textit{m/z} = 30): Formaldehyde (no. 2; LCI = 3; ARM = 206) is identified here unambiguously via M and M-H and it was previously reported for comet 67P \citep{schuhmann2019a}. This molecule was also observed in other comets from the ground \citep{biver2019} and detected in many sources in the ISM \citep{mcguire2022}.
	    \item CH$_4$O (\textit{m/z} = 32): Methanol (no. 3; LCI = 3; ARM = 100), unambiguously identified here via M and M-H, was detected in comet 67P previously \citep{schuhmann2019a}. It was studied from the ground in various comets \citep{biver2019,dellorusso2016} and is commonly detected in the ISM \citep{mcguire2022}.
	    \item C$_2$H$_2$O (\textit{m/z} = 42): Ketene (no. 4; LCI = 3; ARM = 71) is identified here via M with high certainty. The isomers, oxirene and ethynol do not have fragmentation patterns in NIST. Only very recently, oxirene could be produced in methanol-acetaldehyde matrices from isomerization of ketene \citep{wang2023} but it was not yet detected in extraterrestrial environments. Ketene was observed in other comets \citep{biver2019} and in the ISM \citep{mcguire2022}.
	    \item C$_2$H$_4$O (\textit{m/z} = 44): Acetaldehyde (no. 5; LCI = 3; ARM = 150), identified here via M and M-H, has been reported for comet 67P previously \citep{schuhmann2019a}. Oxirane -- detected in the ISM as reviewed by \citet{mcguire2022} -- produces less intensity on \textit{m/z} = 43 and, hence, is a less favorable candidate. However, we cannot completely rule out a marginal presence at the expense of some acetaldehyde. Ethenol does not have fragmentation data in NIST.
	    \item C$_2$H$_6$O (\textit{m/z} = 46): Dimethyl ether (no. 6; LCI = 3; ARM = 3.1) and ethanol (no. 7; LCI = 3; ARM = 4.1) both must be present in 67P's coma with high certainty to explain the observed sum fragmentation pattern. Acetone (no. 10) is contributing to the same group of signals, too, see below. Ethanol produces signals on M, M-H, and M-CH$_3$ (i.e., CH$_3$O on \textit{m/z} = 31) while dimethyl ether shows on M and M-H. Ethanol has been identified in comet 67P previously \citep{schuhmann2019a} and has been reported for other comets \citep{biver2019} and the ISM \citep{mcguire2022}. While in the ISM dimethyl ether has been convincingly identified \citep{mcguire2022}, for comets only an upper limit value is available \citep{biver2019}.
	    \item C$_3$H$_2$O (\textit{m/z} = 54): 2-Propynal (no. 8; LCI = 2; ARM = 1.7) does not match the observed M/M-H ratio well, which makes this isomer less likely as sole isomer contributing to the observed sum fragmentation pattern analyzed in this work. However, other isomers do not have fragmentation data in NIST and, hence, cannot be included in this analysis. Propynal has been reported in the ISM \citep{mcguire2022}.
	    \item C$_3$H$_4$O (\textit{m/z} = 56): 2-Propenal (no. 9; LCI = 2; ARM = 29) is identified via M and M-H signals. But M-H cannot be explained solely by 2-propenal and needs contributions from other species with higher masses. In addition, structural isomers of 2-propenal might contribute. However, other isomers do not have fragmentation data in NIST and, hence, cannot be included in this analysis. 2-Propenal has been reported in the ISM \citep{mcguire2022}.
	    \item C$_3$H$_6$O; \textit{m/z} = 58: Acetone (no. 10; LCI = 3; ARM = 118) and propanal (no. 11; LCI = 3; ARM = 48) are both needed to explain the sum fragmentation pattern investigated in this work -- propanal to explain the observed signal on \textit{m/z} = 57 and acetone to explain that on \textit{m/z} = 43. Propanal or acetone have been identified in 67P previously \citep{schuhmann2019a}; however, the two structural isomers could not be distinguished. Both these isomers have been detected in the ISM \citep{mcguire2022}, but only acetone has been observed in comets from the ground to date \citep{biver2019}. The ISM hosts yet another structural isomer, namely propylene oxide also known as 2-methyloxirane \citep{mcguire2022}. Propylene oxide may or may not be marginally present at the expense of some acetone and propanal in comet 67P. Moreover, it is almost indistinguishable from methoxyethene based on its fragmentation pattern. Notably, no C$_n$H$_m$O species heavier than 58 Da have been firmly detected in the ISM as of mid-2021, which is the cut-off time of the census by \citet{mcguire2022}.
	    \item C$_3$H$_8$O (\textit{m/z} = 60): Isopropanol (no. 12; LCI = 2; ARM = 19) and \textit{normal}-propanol or \textit{n}-propanol (no. 13; LCI = 3; ARM = 5.0) both must be present to explain the observed sum signals appropriately. However, the fragmentation pattern of methoxyethane is similar to that of isopropanol and its presence cannot be excluded. Methoxyethane has recently been proposed in the framework of a re-analysis of COSAC data collected from 67P's surface dust \citep{leseigneur2022}, see details below. Propanol has not been observed in any comet previously but in the galactic center source Sagittarius B2 (Sgr B2), for which \citet{belloche2022} reported the presence of both the \textit{normal} and the \textit{iso} species.
	    \item C$_4$H$_4$O (\textit{m/z} = 68): Here, furan (no. 14; LCI = 2; ARM = 1.7) is identified with certainty via M as most other isomers yield M and M-H (where no signal is observed). Only 3-butyn-2-one could partially contribute. However, M-CH$_3$ (\textit{m/z} = 53) strongly limits its contribution. Furan was searched for in various sources in the ISM but to date only upper limits were reported, see discussion below. \citet{haenni2022} already mentioned the presence of furan based on the data set evaluated here, but no other detections in comets are known to date.
	    \item C$_4$H$_6$O (\textit{m/z} = 70): 2,3-Dihydrofuran (no. 15; LCI = 2; ARM = 8.3) was identified with high certainty via M and M-H. Most likely, both isomers of dihydrofuran contribute, 2,3-dihydrofuran and 2,5-dihydrofuran. Additional contributions of other isomers like 2-butenal, cyclopropanecarboxaldehyde, or (Z)-1,3-butadien-1-ol cannot be fully ruled out because they have similar fragmentation patterns. In most cases some of the fragments are limiting, though. \citet{haenni2022} already reported the presence of dihydrofuran based on the same data set evaluated here, but no other detections in comets are known to date.
	    \item C$_4$H$_8$O (\textit{m/z} = 72): Butanal (no. 16; LCI = 3; ARM = 12) and tetrahydrofuran (THF; no. 17; LCI = 3; ARM = 64) are both present with high certainty to explain the observed signals. THF yields M and M-H while butanal yields M but no M-H. Ethyloxirane could replace them but the abundance would have to be very high and lead to an overshooting M-CH$_3$ fragment on \textit{m/z} = 57. Moreover, in that case oxirane should probably be dominant too. Butanal, or isomers of it, has been identified in 67P previously \citep{schuhmann2019a}. THF has already been mentioned in \citet{haenni2022}. Despite the furan group of molecules can explain well the peak group \textit{m/z} = 68--72, minor contributions of other structural isomers cannot be ruled out and errors on the estimated relative abundances are considerable.
	    \item C$_4$H$_{10}$O (\textit{m/z} = 74): 2-Butanol (no. 18; LCI = 2; ARM = 8.3) is identified here mainly via its M-CH$_3$ fragment (M is very small) and is probably present. 1-Butanol may be contributing as well. However, it cannot be identified with high certainty and, if present, its contribution is limited.
	    \item C$_5$H$_4$O (\textit{m/z} = 80): 2,4-Cyclopentadiene-1-one (no. 19; LCI = 2; ARM = ?) is the only molecule with this sum formula registered in NIST but it does not have fragmentation data. However, a cyclic ketone is expected to produce a strong M signal (and likely an M-H signal too), which we assume here. We cannot rule out that the other isomers not listed in NIST may produce M and M-H signals in an appropriate ratio and hence might be more suitable. Due to the missing fragmentation pattern of 2,4-cyclopentadiene-1-one, we cannot deduce a relative abundance. However, since cyclopentadiene is a relatively abundant molecule among 67P's complex organics (cf. \citet{haenni2022}), we consider 2,4-cyclopentadiene-1-one a probable species.
	    \item C$_5$H$_6$O (\textit{m/z} = 82): 2-Methylfuran (no. 20; LCI = 3; ARM = 11) is identified here with high certainty via M and M-H. The position of the methyl group can be 2 or 3 and is most probably a mixture of both. Other isomers with fragmentation data in NIST do not produce M-H. Unfortunately, for 2H- and 4H-pyran no fragmentation pattern is available, but they are expected to be present at least in low abundance in analogy to the furan group of molecules.
	    \item C$_5$H$_8$O (\textit{m/z} = 84): We identify 2,3-dihydro-4-methylfuran (no. 21; LCI = 3; ARM = 20) via M and M-H signals with high certainty. Other isomers with fragmentation data available in NIST do not produce significant M-H signals and, hence, are less likely according to Occam's razor.
	    \item C$_5$H$_{10}$O (\textit{m/z} = 86): Tetrahydropyran (no. 22; LCI = 3; ARM = 25) is identified via M and M-H signals with high certainty. Other isomers with available fragmentation data in NIST do not produce comparable M-H signals and, hence, are less likely candidates based on Occam's razor.
	    \item C$_5$H$_{12}$O (\textit{m/z} = 88): 1-Ethoxypropane (no. 23; LCI = 3; ARM = 3.9) and 2-methoxy-2-methylpropane (no. 24; LCI = 3; ARM = 0.2) are both needed to explain M (1-ethoxypropane) and M-H (2-methoxy-2-methylpropane). No isomer with fragmentation data available in NIST explains M and M-H simultaneously.
	    \item C$_6$H$_6$O (\textit{m/z} = 94): Phenol (no. 25; LCI = 2; ARM = 7.7) is here identified via strong M. Despite that vinyl-furan has a very similar fragmentation pattern, a substantial contribution of this molecule is probably not likely because it would have to be more abundant than furan itself. However, \citet{haenni2022} reported considerable abundances of benzene, which makes phenol a reasonable candidate species. It has to be noted, though, that many isomers do not have fragmentation data in NIST and it is well possible that one of them may better fit the observed sum fragmentation pattern or at least contribute to it.
	    \item C$_6$H$_8$O (\textit{m/z} = 96): 2,5-Dimethylfuran (no. 26; LCI = 3; ARM = 11) can be identified with high certainty via M, M-H, and M-CH$_3$. However, the positions of the two methyl groups are not clearly defined and contribution of isomers with other positions or an isomer mixture is possible.
	    \item C$_6$H$_{10}$O (\textit{m/z} = 98): 3-Methoxycyclopentene (no. 27; LCI = 3; ARM = 33) is identified via M, M-H, and M-CH$_3$ with high certainty. 3-Methoxycyclopentene is the only available isomer with a strong M-H signal and a reasonable candidate molecule, given the relatively high abundance of cyclopentene \citep{haenni2022}.
	    \item C$_7$H$_6$O (\textit{m/z} = 106): Benzaldehyde (no. 28; LCI = 3; ARM = 4.1) is identified via strong M and M-H signals with high certainty and was already mentioned in \citet{haenni2022}. Isomers with fragmentation data available in NIST do not produce M-H and, hence, are less likely candidates.
	    \item C$_7$H$_8$O (\textit{m/z} = 108):  Benzyl alcohol (no. 29; LCI = 1; ARM = 4.1), identified here via strong M and M-H in the suitable ratio, was already mentioned in \citet{haenni2022}. Most likely, benzyl alcohol is part of a mixture of different structural isomers based on a substituted benzene ring, as these molecules show similar fragmentation patterns. Especially the presence of anisole might add to the observed intensity on \textit{m/z} = 108, which cannot be explained by benzyl alcohol alone.
	    \item C$_7$H$_{10}$O (\textit{m/z} = 110): 2,3,5-Trimethylfuran (no. 30; LCI = 2; ARM = 9.3) is identified via strong M and M-H signals in a suitable ratio. The presence of this molecule is probable because other isomers with fragmentation data available from NIST do not match the observed data as well.
	    \item C$_7$H$_{14}$O (\textit{m/z} = 114): 2-Methylcyclohexanol (no. 31; LCI = 1; ARM = 30) is identified via strong M and absent M-H. However, several other structural isomers expose a similar fragmentation pattern and, hence, the identification comes with a low level of certainty.
	    \item C$_8$H$_6$O (\textit{m/z} = 118): We can identify benzofuran (no. 32; LCI = 3; ARM = 1.9) via strong M and absent M-H signals. As other structural isomers with fragmentation data available from NIST do not match well with the observed data, we consider benzofuran to be a very likely candidate.
	    \item C$_8$H$_8$O (\textit{m/z} = 120): 4-Methylbenzaldehyde (no. 33; LCI = 2; ARM = 4.3) is identified via strong M and M-H signals. The position of the methyl group can be 2, 3, or 4 and cannot be distinguished in the framework of this analysis because the fragmentation patterns of the different isomers are too similar. A mixture of multiple isomers is likely present.
	    \item C$_8$H$_{10}$O (\textit{m/z} = 122): Ethoxybenzene (no. 34; LCI = 1; ARM = 19) is identified via strong M. However, the identification has a low certainty because other isomers with similar fragmentation patterns could and likely do contribute.
	    \item C$_8$H$_{12}$O (\textit{m/z} = 124): We identify 6-methyl-3,5-heptadien-2-one (no. 35; LCI = 1; ARM = 8.3) via M and M-CH$_3$. Via M-CH$_3$ fragment, this molecule contributes to the underpopulated \textit{m/z} = 109 measured signal. However, some of the other isomers might lose a methyl group too and thus may have a similar effect. Hence, the identification has a low certainty.
	    \item C$_8$H$_{14}$O (\textit{m/z} = 126): 2,6-Dimethylcyclohexanone (no. 36; LCI = 1; ARM = 19) is a possible candidate identified via strong M. But some other cyclic isomers may and likely do contribute to the observed data.
   \end{itemize}

\subsection{C$_n$H$_m$O$_2$ species}\label{subsec:CHO2}
	
   \begin{itemize}	
	    \item CO$_2$ (\textit{m/z} = 44): This work identifies carbon dioxide (no. 51; LCI = 3; ARM = 1071) via its strong M and M-O (i.e., CO on \textit{m/z} = 28) signals. Like carbon monoxide, carbon dioxide has been identified in comet 67P previously \citep{leroy2015,migliorini2016,fink2016,rubin2019b} and was studied in great spatial and temporal detail \citep{combi2020,laeuter2020}. CO$_2$ is not observable via its rotational spectrum from the ground as the molecule is missing a permanent dipole moment. \citet{snodgrass2017} imaged 67P's CO$_2$ coma in the infrared in their Figure 8. An overview of CO$_2$ measurements for other comets can be found, e.g., in \citet{ahearn2012}. In the ISM, CO$_2$ ice was detected in absorption \citep{mcguire2022}. 
	    \item CH$_2$O$_2$ (\textit{m/z} = 46): Formic acid (no. 52; LCI = 3; ARM = 36), identified here with great certainty via M and M-H, was previously reported in comet 67P \citep{schuhmann2019a}. It was also detected in various other comets \citep{biver2019} and the ISM \citep{mcguire2022}.
	    \item C$_2$H$_2$O$_2$ (\textit{m/z} = 58): Glyoxal (no. 53; LCI = 2; ARM = 2.5), also known as glyoxal, is possibly present, identified via strong M, but the derived abundance is probably wrong. This is because the fragmentation pattern in NIST seems to be faulty. Specifically, no combination of atoms can explain the strong fragment signal on \textit{m/z} = 31, which represents almost 90\% of the intensity of the main signal on \textit{m/z} = 29 (CHO cation). We hypothesize that the reference sample might have been contaminated with water-adduct or multimer species. The structural isomers, acetylenediol and acetolactone, however, do not have fragmentation data in NIST.
	    \item C$_2$H$_4$O$_2$ (\textit{m/z} = 60): Acetic acid (no. 54; LCI = 3; ARM = 11), glycolaldehyde (no. 55; LCI = 3; ARM = 44), and methyl formate (no. 56; LCI = 3; ARM = 44) are very likely present simultaneously. Acetic acid is clearly identified via strong M, M-CH$_3$, and M-OH and has been previously reported to be present in comet 67P by \citet{schuhmann2019a}. A combination of acetic acid and methyl formate and/or glycolaldehyde (identified via strong M) is necessary to explain M appropriately while not overshooting M-CH$_3$. Both methyl formate and glycolaldehyde have been identified in comets from the ground \citep{biver2019}, glycolaldehyde quite recently in comet C/2014 Q2 Lovejoy \citep{biver2015}. However, based on the mass spectrometric data analyzed here, their abundances are not well constrained. This is because their fragmentation patterns are very similar and we have artificially fixed their ratio to 1. Notably, the glycoladehyde:methyl formate ratio is about 0.2 in comet C/2014 Q2 Lovejoy \citep{biver2019}.
	    \item C$_2$H$_6$O$_2$ (\textit{m/z} = 62): Ethylene glycol (no. 57; LCI = 3; ARM = 50) is identified with high certainty via M and M-H. A limited contribution of the structural isomer dimethyl peroxide is possible. But dimethyl peroxide alone cannot sufficiently explain M-H. Ethylene glycol has been identified in comet 67P previously \citep{schuhmann2019a} and was reported for other comets \citep{biver2019} and the ISM \citep{mcguire2022}. 
	    \item C$_3$H$_4$O$_2$ (\textit{m/z} = 72): Here, we identify 2-propenoic acid (no. 58; LCI = 3; ARM = 6.3) via strong M. Other isomers with fragmentation data available from NIST do not have strong M signals and are thus less likely.
	    \item C$_3$H$_6$O$_2$ (\textit{m/z} = 74): Propanoic acid (no. 59; LCI = 3; ARM = 39), here clearly identified via M and M-H, was already mentioned by \citet{haenni2022}. Moreover, methyl acetate (or isomers of it) was reported to be present in comet 67P \citep{schuhmann2019a}. In the ISM, ethyl formate, methyl acetate, and hydroxy acetone were detected \citep{mcguire2022}. However, based on the data evaluated here, all three of these isomers are less favorable candidates than propanoic acid because they produce little to no M and no M-H signals. If present, their abundances cannot be constrained well by mass spectrometry alone.
	    \item C$_3$H$_8$O$_2$ (\textit{m/z} = 76): 2-Methoxyethanol (no. 60; LCI = 3; ARM = 27), methylal (no. 61; LCI = 2; ARM = 4.0), and 1,2-propanediol (no. 62; LCI = 2; ARM = 24) are most likely present simultaneously. The combination of 2-methoxyethanol (yields M and a strong C$_2$H$_5$O fragment on \textit{m/z} = 45), 1,2-propanediol (yields M-CH$_3$ but no relevant M or M-H signals), and methylal (yields M-H) provides a sufficient explanation of the observed data. No other isomers with fragmentation data in NIST apart from 1,2-propanediol could contribute substantially to the observed signal on \textit{m/z} = 61 and only 2-methoxyethanol yields relevant M. Consequently, the presence of both these molecules is considered likely -- 1,2-propanediol being a bit less likely than 2-methoxyethanol because M signal can be used for the identification. Methylal is identified with less confidence too, as other isomers could yield M-H signals but for methylal the relative intensity of M-H is greatest.
	    \item C$_4$H$_4$O$_2$ (\textit{m/z} = 84): 2(3H)-Furanone (no. 63; LCI = 2; ARM = 12) is identified via strong M signal. For the position of the reduced C atom in the furan core structure, 3H or 5H are equally possible. Also, a relevant contribution of the structural isomer 2-butynoic acid cannot be ruled out based on the similarity of the reference fragmentation patterns. Unfortunately, the structural isomers 1,4-dioxine and 1,2-dioxine do not have fragmentation data in NIST. But in analogy to the furan and pyran groups of molecules, dioxines (more likely the non-peroxide isomer) may be present too given that dioxanes are likely present (cf. no. 66 and no. 67 below).
	    \item C$_4$H$_6$O$_2$ (\textit{m/z} = 86): Cyclopropanecarboxylic acid (no. 64; LCI = 1; ARM = 12) is likely contributing to M-H, while $\gamma$-butyrolactone (no. 65; LCI = 1; ARM = 66) is explaining the observed M signal (and contributes to M-H). As cyclopropanecarboxylic acid does not have a relevant M signal, the reference data was normalized to M-H on \textit{m/z} = 85. Other combinations of isomers are possible but slightly less likely and many isomers with fragmentation data in NIST do not yield significant M-H.
	    \item C$_4$H$_8$O$_2$ (\textit{m/z} = 88): The combination of 1,3-dioxane (no. 66; LCI = 3; ARM = 6.2; strong M-H but no M) and 1,4-dioxane (no. 67; LCI = 3; ARM = 13; strong M but no M-H) explains the observed data well. 1,2-Dioxane is a peroxide and a less stable structural isomer with no fragmentation data available from NIST. However, peroxides are observed in comet 67P (see diethyl peroxide, no. 70, below) and we cannot rule out its contribution here. Butanoic acid (no. 68; LCI = 2; ARM = 13) is needed in addition to the dioxane isomers to explain M-CH$_3$ on \textit{m/z} = 73 and is thus considered to be likely present as well. 
	    \item C$_4$H$_{10}$O$_2$ (\textit{m/z} = 90): 1-Ethoxy-1-methoxyethane (no. 69; LCI = 2; ARM = 19) does not yield relevant M and is identified via M-H (normalization to M-H on \textit{m/z} = 89) while diethyl peroxide (no. 70; LCI = 2; ARM = 3.8) is identified via strong M and does not yield significant M-H. Together they explain the observed data well. We recall that also dimethyl peroxide may be present (see C$_2$H$_6$O$_2$ above). Methyl ethyl peroxide is not listed in NIST and hence cannot be clearly identified or ruled out.
	    \item C$_5$H$_4$O$_2$ (\textit{m/z} = 96): 3-Furaldehyde (no. 71; LCI = 1; ARM = 2.2) is identified in this work via M and M-H signals. The structural isomer furfural exposes a very similar fragmentation pattern and is a possible alternative candidate.
	    \item C$_5$H$_6$O$_2$ (\textit{m/z} = 98): We identify 2-furanmethanol (no. 72; LCI = 3; ARM = 18) with high certainty via its strong M and M-H signals. However, the position of the hydroxy group could be either 2 or 3 and is most likely a mixture of both.
	    \item C$_5$H$_8$O$_2$ (\textit{m/z} = 100): Cyclopropanecarboxylic acid methyl ester (no. 73; LCI = 2; ARM = 2.5) is identified via M and M-H. It is the only isomer with fragmentation data in NIST that yields significant M-H. Moreover, cyclopropanecarboxylic acid (no. 64 above) is likely present as well.
	    \item C$_5$H$_{10}$O$_2$ (\textit{m/z} = 102): While propanoic acid ethyl ester (no. 74; LCI = 2; ARM = 15) is identified via M, 4-methyl-1,3-dioxane (no. 75; LCI = 2; ARM = 20) yields M and M-CH$_3$. However, due to absence of significant M, the latter fragmentation pattern had to be normalized to the M-H signal. The methyl group is more likely to be on position 4 than on position 2 because position 2 would yield too much intensity on M-CH$_3$ as compared to M-H. However, minor contributions of this isomer or other structural isomers without significant M and M-H signals cannot be ruled out. Pentanoic acid, for instance may or may not be present in very low abundance.
	    \item C$_6$H$_6$O$_2$ (\textit{m/z} = 110): Hydroquinone (no. 76; LCI = 2; ARM = 2.6) is identified via strong M signal but other isomers with the hydroxy group on different positions of the benzene ring are equally possible, i.e., resorcinol also known as 1,3-benzenediol and catechol also know as 1,2-benzenediol. Less likely is 1-(2-furanyl)-ethanone, but a limited contribution especially to better explain the C$_5$H$_3$O$_2$ fragment on \textit{m/z} = 95 is possible.
	    \item C$_7$H$_6$O$_2$ (\textit{m/z} = 122): Benzoic acid (no. 77; LCI = 3; ARM = 25) is identified with high certainty via characteristic M and M-OH signals. This molecule was previously reported in \citep{haenni2022}.
   \end{itemize}

\section{Discussion}\label{sec:disc}
\begin{figure*}
   \centering
   \includegraphics[width=\textwidth]{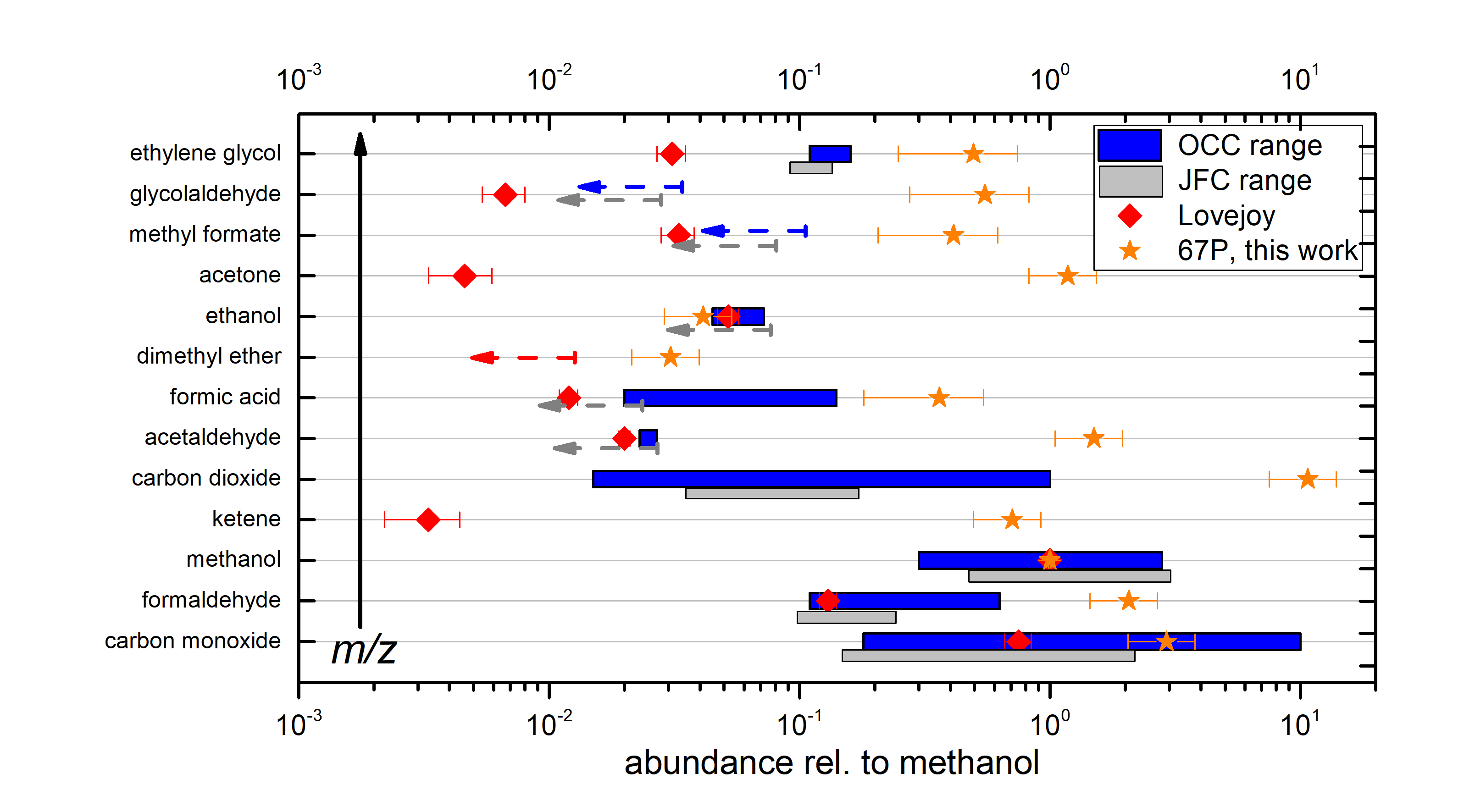}
      \caption{Comparison to O-bearing organic molecules observed in comets remotely as reviewed by \citet{biver2019}. Cometary abundance data from \citet{biver2019} were renormalized from water to methanol. In addition to ranges for both Jupiter Family Comets (JFCs) and Oort Cloud Comets (OCCs), also the values for the long-period OCC C/2014 Q2 (Lovejoy) are included because glycolaldehyde was detected in this comet for the first time \citet{biver2015}. The OCC and JFC ranges for CO$_2$ were taken from \citet{pinto2022} and were normalized with the respective average value of methanol from \citet{dellorusso2016}. Dynamically new comets were not considered. Left-pointing arrows indicate upper limits that have been derived from non-detections of the respective species.}
         \label{fig:comp-com}
   \end{figure*}

In the following, we discuss the different O-bearing molecules identified in the framework of the method applied in this work. Specifically, we compare comet 67P's inventory to other reservoirs of extraterrestrial organics, namely, other comets \citep[][Fig. \ref{fig:comp-com}]{biver2019}, the ISM \citep[][Table \ref{tab:comp-ISM}]{mcguire2022}, as well as SOM extracted from the Murchison meteorite \citep[][Table \ref{tab:comp-met}]{botta2002,sephton2002}. Before discussing the different chemical functional groups of O-bearing organic molecules in the reservoirs mentioned above (in order of decreasing priority of the functional group) while highlighting the relevance to bio-, prebiotic, and astrochemistry, we briefly comment on the reservoirs and the characteristics and caveats of the attempted comparison individually:\\
Fig. \ref{fig:comp-com} compares our findings to ground-based observations of mostly the rotational spectra of O-bearing complex organics in the two comet populations -- Jupiter Family Comets (JFCs; e.g., comet 67P) and Oort Cloud Comets (OCCs; e.g., comet C/2014 Q2 (Lovejoy)) -- as reviewed by \citet{biver2019} and \citet{pinto2022}. CO and CO$_2$ are the major components of cometary ices beside water \citep{mumma2011,biver2019} and they have already been studied in detail in comet 67P previously \citep{laeuter2020,combi2020}. Even around the comet's perihelion, CO and CO$_2$ originate mostly from the icy bulk material \citep{laeuter2019} and, hence, are largely unrelated to the desorption of complex O-bearing organics from ejected cometary grains. As cometary reference species, we included their abundances relative to methanol as observed on 3 August 2015 in Fig. \ref{fig:comp-com}. While the CO abundance lies just outside the upper end of the respective JFC range, the CO$_2$ one exceeds it by almost two orders of magnitude. The corresponding CO/CO$_2$ ratio is 0.27. It is important to note, however, that this is a momentary value and substantial variation of this ratio has been reported \citep{haessig2015}. The variability is not only caused by variations in the outgassing itself but also by changes in the relative position of the spacecraft with respect to the comet and by the comet's rotation. Notably, the methanol/water ratio can vary considerably within just a few hours, too. Except for ethanol, our relative abundance estimates are consistently higher than the other cometary abundances visualized in Fig. \ref{fig:comp-com}, sometimes by more than an order of magnitude. In addition, probably also the specific measurement conditions contribute to the observed deviations. In this work, we analyzed data from a very dusty period close to the comet's perihelion and heavy organics are expected to be released mostly from dust grains. During this period dust was released in short-lived, local dust outbursts \citep{vincent2016}, which means that the dust density and hence the density of heavy organics released from dust was by no means homogeneous on the sunward side of the nucleus. Variable illumination conditions - during perihelion passage emission was dominated by the south-pole region (e.g., \citet{laeuter2019}) - are expected to play a subordinate role. \citet{rubin2019b} have studied the comet's volatile bulk material farther away from the Sun, when there was less sublimation from dust. They find, for instance, that formaldehyde is 152\% (here: 206\%), formic acid 6.2\% (here: 36\%), and acetic acid 1.6\% (here: 11\%) relative to methanol. Values from \citet{rubin2019b} are closer to the values reviewed by \citet{biver2019}.\\
At this point, we also would like to mention a recent re-analysis of data collected by the time-of-flight mass spectrometer COSAC 25 min after initial touchdown of Rosetta's Philae lander unit on 67P's surface. This new study, published by \citet{leseigneur2022} and up-dating previous results from \citet{goesmann2015}, explained the observed overall fragmentation pattern ranging from \textit{m/z} = 12 to 63 with a suite of 12 molecules. These molecules have been identified with a high likelihood based on a least-squares fitting algorithm applied to a set of 120 pre-selected and partially 'hand-picked' reference spectra. Half of them are O-bearing complex organic species, namely, acetaldehyde, acetone, methoxyethane, ethylene glycol, 2-methoxypropane, and cyclopentanol. Acetaldehyde, acetone, and ethylene glycol have been securely identified in this work as well, whereas methoxyethane may be present but cannot be clearly distinguished from isopropanol which shows a very similar fragmentation pattern, hampering unambiguous identification. 2-Methoxypropane does not yield a significant M signal and secure mass-spectrometry-based detection is difficult. Cyclopentanol does yield a minor amount of M but tetrahydropyran with strong M and M-H signals must be favored according to our method of analysis. The COSAC mass spectrum evaluated by \citet{goesmann2015} and \citet{leseigneur2022} was collected in the so-called sniffing mode and it was assumed that, during first contact with the cometary surface, some excavated material entered the exhaust region of the instrument and evaporated there at local temperatures of 12 to 15 °C. Such conditions would be considerably different from those prevalent during the time when the DFMS data investigated in this work was collected, as ejected dust grains can reach temperatures of multiple hundred degrees Celsius in the coma \citep{lien1990} and also heavier species can sublimate.

Table \ref{tab:comp-met} groups the here selected cometary species according to their chemical functionalities in order to compare the abundances of the groups in the two reservoirs, comet 67P (this work) and Murchison SOM (\citet{alexander2017}, updated from \citet{botta2002}, updated from \citet{cronin1988}). The Murchison chondrite is one of the most organic-rich and best-characterized meteorites to date. \citet{haenni2022} have found striking similarities between Murchison SOM as reported by \citet{alexander2017} and comet 67P's complex organics in terms of average sum formula, which, according to these authors, might indicate shared prestellar history. Here, our candidate molecules have been grouped according to priority of the functional groups in cases where two different functionalities are present within the same molecule. This means that, e.g., glycolaldehyde is assigned to the group of aldehydes and not to the group of alcohols, despite its hydroxyl function. We expect that the relative sum abundances of the different groups are less error-prone than the relative individual abundances as uncertainties partially cancel each other out. Despite the fact that some species, like straight chain molecules, are systematically underestimated due to absent or weak M signals, a rough comparison of both these reservoirs of complex organics is still meaningful, given that this bias is likely affecting the different groups of molecules in a similar way.\\
Also the analysis of meteoritic SOM comes with certain biases: First and foremost, small and hence highly volatile molecules are probably mostly lost or have been transformed into more complex molecules during aqueous alteration or metamorphism in the asteroid parent body, long before the meteorite is collected and laboratory analyses are initiated. Compared to organic species in comets, consequently, a shift in the mass-distribution of meteoritic organic molecules towards higher masses is expected. In untargeted high-resolution mass-spectrometric investigations of meteoritic SOM after simple solvent extraction, an extremely high chemical diversity was demonstrated over a broad mass range (some species had masses >400 Da), see, e.g., \citet{schmitt-kopplin2010}. However, usually specific classes of compounds with a certain (pre)biotic or biological interest (such as amino acids, sugars, nucleobases etc.) or molecules for which an established chemical protocol of detection was already known and optimized (carboxylic acids, aldehydes, aromatic compounds etc.) are investigated in targeted analyses. This is a further caveat of SOM investigations as it is obvious that from such studies SOM cannot be characterized comprehensively and in an unbiased manner. Partly related to this is another bias, namely, almost all studies of SOM followed an extraction protocol including an acid hydrolysis step \citep{botta2002}during which esters can be hydrolyzed into an alcohol and a carboxylic acid moiety. Because water as main solvent is found in large excess, the hydrolysis will be total and no ester molecules will be present anymore but rather their hydrolysis products would lead to an overestimation of the carboxylic acid and alcohol budget. Similarly, cyclic esters (lactones) will be detected as hydroxycarboxylic acids under these conditions. Unfortunately, this has not been investigated thoroughly for the case of esters and lactones to the best of our knowledge. However, the bias has been studied and quantified for amino acids (note that amide functions are hydrolyzed into amines and carboxylic acids) by measuring their abundance in solvent extracts with and without hydrolysis \citep{martins2015}. Such studies helped to show that most of those species are not 'free' molecules present in the meteorite bulk, but they are fragments of bigger (but still soluble) molecules. In that sense it is not surprising that molecules with two or even three chemical functions (especially carboxylic acid, amine and/or alcohol functions) are particularly abundant in SOM (targeted) findings. After an acid hydrolysis extraction, this may support the idea that those species are in fact originally fragments (or monomers) bound together in larger macromolecules (or polymers).

\begin{table*}[h!t]
\caption[]{Selected molecules from this work are compared to Murchison SOM$^{(a)}$ in terms of (relative) sum abundances of functional groups.}
\centering 
\begin{tabular}{lllll}
\toprule
Species                   & SOM conc. (ppm)     & SOM rel. to acids   & 67P est. abundance    & 67P rel. to acids     \\
\midrule
Carboxylic acids          & $>$300              & 1                   & 142                   & 1                     \\
Hydroxycarboxylic acids   & 15                  & 0.05                & xxx$^{(b)}$                   & xxx$^{(b)}$                    \\ 
\parbox{4cm}{Dicarboxylic acids and \\hydroxydicarboxylic acids} & 14 & 0.05          & xxx$^{(b)}$                   & xxx$^{(b)}$     \\
Esters linear             & -                   & -                   & 59                    & 0.42                   \\
Esters cyclic/lactones$^{(c)}$ &	-             & -                   & 78                    & 0.55                    \\
Aldehydes                 &	27                  & 0.09                & 516                   & 3.63                    \\
Ketones                   &	                    &                     & 217                   & 1.52                    \\
Alcohols                  &	11                  & 0.04                & 223                   & 1.57                     \\
Polyols                   &	$>$8                & 0.03                & 76                    & 0.53                     \\
Ethers linear             &	-                   & -                   & 83                    & 0.58                     \\
Ethers cyclic$^{(a)}$     &	-                   & -                   & 191                   & 1.34                     \\
Peroxides                 &	-                   & -                   & 4                     & 0.03                     \\
   \bottomrule
		\end{tabular}
	\label{tab:comp-met}
	\tablefoot{$^{(a)}$ Murchison SOM has been reviewed in \citet{botta2002} and \citet{alexander2017} and literature cited therein. Here we consider the values listed in Table 1 in \citet{alexander2017}. $^{(b)}$ Expected M signals of hydroxy-, di-, and hydroxydicarboxylic acids seem to be absent or below DFMS detections limits, which is indicated by (xxx). $^{(c)}$ Cyclic esters and ethers are so-called O-bearing heterocycles.}
\end{table*}

Last but not least, table \ref{tab:comp-ISM} lists the 27 neutral organic O-bearing molecules, composed of four or more atoms (radicals and disputed detections are not considered), which have been identified in the ISM to date. At a small number of exceptions, all molecules can be found in \citet{mcguire2022}. We indicate for each molecule, whether or not it can be detected in the mass-spectrometric data set analyzed for this work and, if yes, list the LCI of our identification. However, the comparison between our in situ mass-spectrometry-based work and ISM detections usually performed with rotational-spectroscopic methods from the ground is subject to methodological biases. Despite the fact that both methods target the gas phase and rely on laboratory reference spectra, mass spectrometry does not always allow for distinction between structural isomers and needs the analyte species to produce a relevant M signal for a secure detection while rotational spectroscopy can access only molecules with a permanent dipole moment. Notably, heteroatoms, like the oxygen atom in the focus of this work, usually introduce a dipole moment to the hydrocarbon molecule and thus help to make it accessible to spectroscopic studies from the ground. A more thorough discussion of the two different techniques and their intrinsic limitations can be found in \citet{haenni2022}.

\begin{table*}[h!t]
\caption[]{List of O-bearing neutral and non-radical organic molecules containing four or more atoms identified in the ISM$^{(a)}$ compared to the findings of this work, including the level-of-confidence indicator (LCI) of our identifications.}
\centering 
\begin{tabular}{lllll}
\toprule
Molecule                  & Constitution formula      & Mass (Da)  & Reference               & Presence in 67P (LCI)   \\
\midrule
Formaldehyde              & H$_2$CO                   & 30         & \citet{mcguire2022}     & yes (3)                 \\
Methanol                  & CH$_3$OH                  & 32         & \citet{mcguire2022}     & yes (3)                 \\
Ketene                    & H$_2$CCO                  & 42         & \citet{mcguire2022}     & yes (3)                 \\ 
Acetaldehyde              & CH$_3$CHO                 & 44         & \citet{mcguire2022}     & yes (3)                 \\
Ethylene oxide            &	\textit{c}-C$_2$H$_4$O    & 44         & \citet{mcguire2022}     & maybe$^{(b)}$           \\
Vinyl alcohol             &	CH$_2$CHOH                & 44         & \citet{mcguire2022}     & $^{(c)}$                \\
Formic acid               & HCOOH                     & 46         & \citet{mcguire2022}     & yes (3)                 \\
Dimethyl ether            & CH$_3$OCH$_3$             & 46         & \citet{mcguire2022}     & yes (3)                 \\
Ethanol                   & CH$_3$CH$_2$OH            & 46         & \citet{mcguire2022}     & yes (3)                 \\ 
Propynal                  & HC$_2$CHO                 & 54         & \citet{mcguire2022}     & yes (2)                 \\
Cyclopropenone            &	\textit{c}-H$_2$C$_3$O    & 54         & \citet{mcguire2022}     & $^{(c)}$                \\
Propenal                  &	CH$_2$CHCHO               & 56         & \citet{mcguire2022}     & yes (2)                 \\
1-Propen-1-on             &	CH$_3$CHCO                & 56         & \citet{fuentetaja2023}  & $^{(c)}$                \\
Acetone                   & CH$_3$COCH$_3$            & 58         & \citet{mcguire2022}     & yes (3)                 \\
Propanal                  & CH$_3$CHCH$_2$O           & 58         & \citet{mcguire2022}     & yes (3)                 \\
Propylene oxide           & CH$_3$CH$_2$OH            & 58         & \citet{mcguire2022}     & maybe$^{(b)}$           \\ 
Isopropanol               &	\textit{i}-C$_3$H$_7$OH   & 60         & \citet{belloche2022}    & yes (2)                 \\
\textit{n}-Propanol       &	\textit{n}-C$_3$H$_7$OH   & 60         & \citet{belloche2022}    & yes (3)                 \\
Methyl formate            & HCOOCH$_3$                & 60         & \citet{mcguire2022}     & yes (3)                 \\
Acetic acid               &	CH$_3$COOH                & 60         & \citet{mcguire2022}     & yes (3)                 \\
Glycolaldehyde            &	CH$_2$OHCHO               & 60         & \citet{mcguire2022}     & yes (3)                 \\
Z-1,2-Ethenediol          & HOCHCHOH                  & 60         & \citet{rivilla2022}     & $^{(c)}$                \\
Ethylene glycol           & HOCH$_2$CH$_2$OH          & 62         & \citet{mcguire2022}     & yes (3)                 \\
3-Hydroxypropenal$^{(d)}$ & OHCHCHCHO                 & 72         & \citet{coutens2022}     & $^{(c)}$                \\
Ethyl formate             & C$_2$H$_5$OCHO            & 74         & \citet{mcguire2022}     & maybe$^{(d)}$           \\
Methyl acetate            & CH$_3$COOCH$_3$           & 74         & \citet{mcguire2022}     & maybe$^{(e)}$           \\ 
Hydroxyacetone            & CH$_3$COCH$_2$OH          & 74         & \citet{mcguire2022}     & maybe$^{(e)}$           \\
   \bottomrule
		\end{tabular}
	\label{tab:comp-ISM}
	\tablefoot{$^{(a)}$ Original literature for ISM species can be found in the 2021-census of \citet{mcguire2022} and only more recent detections are referenced specifically. Disputed detections are not considered. $^{(b)}$ May or may not be present at the expense of the structural isomer(s). $^{(c)}$ No reference data available from NIST. $^{(d)}$ Identification tentative. $^{(e)}$ Molecule may or may not be present with low relative abundance but the fragmentation pattern does not show any significant M or M-H signals for a clear identification.}
\end{table*}

\subsection{Carboxylic acids (R--COOH) and carboxylate esters (R--COO--R)}\label{subsec:acids}
Carboxylic acids are of biological interest, as especially the linear ones among them could form membranes and vesicles via hydrophobic interactions. Many of them have been identified to be part of meteoritic SOM, cf. \citet{botta2002} and literature referenced therein and our Table \ref{tab:comp-met}. The latter shows that Murchison SOM seems to be rich in carboxylic acids compared to comet 67P's volatile organics, where carboxylic acids are clearly not the dominant O-bearing functional group. As discussed above, the extraction protocol can introduce a severe bias for the quantification of carboxylic acids as well as esters and amides as the latter two may by hydrolyzed into acids depending on the extraction conditions. This may lead to an overestimation of the abundance of carboxylic acids in meteoritic SOM. In Murchison, \citet{yuen1984} reported acetic acid to be the most abundant monocarboxylic acid. From our data, the estimation of the effective total abundances of carboxylic acids is difficult as well because we know that they may be trapped in the form of ammonium salts \citep{altwegg2020,poch2020,altwegg2022}. However, in the data from 3 August 2015 analyzed for this work, salts are not considered a major contributor. For the case of acetic acid, the coexistence of the two other isomers, namely, methyl formate and glycolaldehyde, introduces an additional factor of uncertainty. We also find clear indications for butanoic acid, while we cannot confirm pentanoic and higher carboxylic acids due to absent M signal. In return, the abundance of propanoic acid might be overestimated because the O$_2$-bearing fragment on \textit{m/z} = 73 (C$_3$H$_5$O$_2$), which is as well the M-H fragment of propanoic acid, is characteristic to exactly those linear carboxylic acids with at least three carbon atoms that we cannot identify. Notably, the branched species are likely to contribute but they expose very similar fragmentation patterns in terms of O-retaining fragments and, hence, cannot be distinguished clearly from the unbranched versions. From the ground, only formic acid has been observed in comets \citep{biver2019}. Unlike in Murchison SOM, in our data set no signals of species with three or more O atoms are identified that could be indicative for hydroxy-carboxylic acids or dicarboxylic acids. The fact that no hydroxy-carboxylic acids seem to be present in comets might find an explanation in the aqueous alteration step in meteoritic parent bodies. It is thought that $\alpha$-hydroxy carboxylic acids are produced during parent body aqueous alteration alongside $\alpha$-amino acids via Strecker-cyanohydrin synthesis from carbonyl compounds and hydrogen cyanide and in presence of ammonia \citep{peltzer1984}. This is a hypothesis, however, that still needs to be supported by systematic SOM analyses from series of chondrites with different alteration degrees. Given the detection after acid hydrolysis, it is possible that these classes of compounds may, at least partly, be fragments of bigger molecules containing ester and amide moieties. In the ISM, acetic acid is the largest carboxylic acid detected \citep{mehringer1997}. In our data, the largest and conclusively identified carboxylic acid is benzoic acid, already mentioned in \citet{haenni2022}.\\
In terms of esters, we state the following: Apart from methyl formate, the abundance of which is not well constrained from our data due to structural isomerism with glycolaldehyde and acetic acid, two lactones are likely present and make a relevant contribution, namely, furanone and its hydrogenated version $\gamma$-butyrolactone. The latter seems to be more abundant than the dehydrogenated furanone. However, relative abundance estimates come with a high uncertainty and it is possible that they are generally overestimated because in both cases structural isomers with similar fragmentation patterns exist and could contribute. From the ground-based observation of other comets, only upper limits for methyl formate were obtained in OCCs other than C/1995 O1 (Hale-Bopp) and C/2014 Q2 (Lovejoy) and in JFCs \citep{biver2019}. In the ISM both methyl and ethyl formate were detected as well as methyl acetate \citep{mcguire2022}. For the Murchison meteoritic SOM, esters have not been quantified to date to the best of our knowledge. As discussed in detail above, a reason may be conversion of esters into carboxylic acids under acid hydrolysis conditions. It might be worth trying to detect and quantify these compounds in chondritic SOM without prior hydrolysis and preferentially by using a solvent extraction protocol that does not involve water but for instance methanol.

\subsection{Aldehydes (R--CHO) and ketones (R--CO--R)}\label{subsec:ald-ket}
Aldehydes and ketones are commonly found in meteorites and may also play an important role, e.g., in the Strecker synthesis of amino acids \citep{botta2002} and possibly in the abiotic synthesis of sugars. It is hypothesized that sugars have been prebiotically formed from formaldehyde via formose reaction either on the early Earth or even in parent bodies during aqueous alteration, see, e.g., \citet{paschek2022} and literature referenced therein. The simplest sugar is the aldose glyceraldehyde (C$_3$H$_6$O$_3$) and its ketose isomer dihydroxyactone. Although searched for in the ISM, glyceraldehyde has not been found yet \citep{hollis2004} and the detection of dihydroxyacetone in the galactic center source Sgr B2 by \citet{widicus-weaver2005} has been disputed \citep{apponi2006}. However, the related molecule glycolaldehyde (C$_2$H$_4$O$_2$) has been detected in the ISM \citep{hollis2000} as well as in comets \citep{biver2015} and will be discussed in more detail below. Only recently, several sugars, including the pentose sugar ribose (C$_5$H$_{10}$O$_5$), have been identified in chondrites via gas chromatography coupled mass spectrometry after acid hydrolysis \citep{furukawa2019}. In its heterocyclic furanose form, ribose is present in the bio-macromolecules ribonucleic acid (RNA) and deoxyribonucleic acid (DNA). For this reason, recurrent efforts were made towards the identification of O-bearing heterocycles. From a chemical point of view, O-bearing heterocycles are mostly cyclic ethers. We discuss further details below under \ref{subsec:eth}.\\
For Murchison SOM, aldehydes and ketones were found to make up a total of 27 ppm, which is roughly an order of magnitude less than the lower limit reported for carboxylic acids \citep{botta2002}. For aldehydes, all possible structural isomers through C4 seem to be present with decreasing concentrations with increasing number of carbon atoms, cf. \citet{jungclaus1976}. A similar case is reported ibidem for ketones. In addition, some higher molecular weight hydrocarbons with polar ketone functionalities, e.g., ketone derivatives of simple polycyclic aromatic hydrocarbons (PAHs), have been tentatively identified \citep{krishnamurthy1992}. From our data, we find a total of aldehydes and ketones that is clearly outweighing the identified carboxylic acids by about a factor of 5, cf. Table \ref{tab:comp-met}. However, it seems that carboxylic acids, rather than aldehydes and ketones, tend to not yield significant M signals, which might lead to an overestimation of this factor.\\
Formaldehyde and acetaldehyde are by far the two most abundant aldehydes identified from our data, followed by propanal and propenal. The relative abundance of glycolaldehyde is poorly constrained due to structural isomerism. In other comets, the following aldehydes have been observed from the ground: formaldehyde, acetaldehyde, and glycolaldehyde \citep{biver2019}. We observe more formaldehyde as compared to methanol, while an inverse ratio is derived from the remote observational data for both OCCs and JFCs \citep{biver2019}, cf. Fig. \ref{fig:comp-com}. Whether or not this might indicate a local and/or temporal dust source \citep{cottin2004,cordiner2014} cannot be determined from this single data set collected over a time period of just a few hours and at one specific cometocentric distance. Therefore, further investigation is necessary. In the ISM, the aldehydes propynal, propenal, and propanal have been listed as detected by \citet{mcguire2022} in addition to the species that have been observed in comets. Formaldehyde, which seems to be the most abundant aldehyde, both in comets and the ISM, has been detected in over a dozen of interstellar sources already in 1969 \citep{snyder1969}.\\
Among the ketones, we find that acetone is very abundant as well as ketene (International Union of Pure and Applied Chemistry (IUPAC) standard name: ethenone). But there seems to be a small number of even larger ketones, as large as dimethylcyclohexanone. Both acetone and ketene have been identified also from ground in comets other than 67P \citep{biver2019}. In the ISM, the ketones cyclopropenone and hydroxyacetone have been reported in addition \citep{mcguire2022}. While for cyclopropenone no reference mass spectrum is available from NIST, hydroxyacetone does not yield significant M and M-H signals. Consequently, even if present, we cannot clearly identify these ketones from our data.

\subsection{Alcohols (R--OH)}\label{subsec:alc}
In biomolecules, alcohols are mostly found in sugars but also in some amino acids and other molecules with biological relevance. The alcohols detected to date in the ISM \citep{mcguire2022} and in comets other than 67P \citep{biver2019} are methanol, ethanol, and ethanediol. In addition, vinyl alcohol has been reported for the ISM, cf. \citet{mcguire2022} and literature referenced therein. \citet{rivilla2022} identified Z-1,2-ethenediol, and \citet{belloche2022} detected \textit{n}- and isopropanol. From our analysis, we find that methanol is abundant while the higher homologous series alcohols (ethanol, propanol, etc.) are not. This is consistent with reports from analysis of Murchison SOM \citep{jungclaus1976}, where methanol is present in absolute abundance of 5 $\mu$g/g and abundances gradually decrease towards the butyl alcohols, which are the most complex alcohols found in that study. While isopropanol could be identified clearly, the identification of \textit{n}-propanol was only tentative. If correct, \textit{n}-propanol seemed to be slightly more abundant than isopropanol based on the gas chromatogram of the head space above the liquid extract. Our data indicate isopropanol to be more abundant than \textit{n}-propanol by a factor of about 4 (subject to substantial uncertainty due to the possible presence of methoxyethane), while a recent study of the galactic center source Sgr B2(N) by \citet{belloche2022} demonstrated a slight dominance of the normal over the iso-version. Modeling work accompanying that latter observation showed that the observed ratio might be a direct inheritance of the branching ratio of the OH radical addition to propylene in dust-grain ice mantles (driven by water photodissociation). However, results from modeling the branched carbon chain chemistry in Sgr B2(N) indicate that the degree of branching rises with molecular size \citep{garrod2017}. Apart from 2-butanol in this study and butyl alcohols in Murchison SOM \citep{jungclaus1976}, no higher homologous series alcohols can be identified. Admittedly, their mass spectrometry-based identification is hampered by the fact that especially larger and unbranched alcohols do not yield significant M signals, cf. Methods section. Nevertheless, it is possible to rule out the presence of large amounts of unidentified alcohols in 67P's coma on the basis of the resonance-stabilized CH$_3$O ion on \textit{m/z} = 31, which is a common fragment of mostly primary alcohols. Our Occam's razor-based solution in Fig. \ref{fig:CHO} does explain well the corresponding measured intensity and additional contributions are limited by the estimated uncertainty of the signal of 30\%. Beyond the homologous series alcohols, also phenol and benzylacohol are likely present in 67P's coma. The latter has been reported already \citep{haenni2022}. The abundances of the even larger alcohols (2-methylcyclohexanol, 2-methoxyethanol, and 2-furanmethanol) may be overestimated as it is unlikely that among the many structural isomers of those molecules only one exists. A possible reason could be that our Occam's razor approach is less appropriate for more complex molecules for which it is more likely that multiple low-abundant isomers contribute to the observed sum fragmentation spectrum, cf. Methods section. Both in Murchison SOM as well as in our data, the sum abundance of aldehydes plus ketones outweighs that of alcohols. From Table \ref{tab:comp-met}, the abundance ratio of alcohols with respect to aldehydes plus ketones is roughly 0.7 for Murchison while for our data it seems to be lower, about 0.4. However, if the underestimation of aliphatic species (see Methods section) is different for the different functional groups, the factor we estimate might be influenced. If, e.g., aliphatic alcohols tend to be underestimated rather than aliphatic ketones and aldehydes, then the effective ratio for comet 67P would be slightly underestimated and thus closer to the Murchison SOM value. Comet C/2014 Q2 (Lovejoy) shows a ratio of roughly 6.5 based on Fig. \ref{fig:comp-com} and the data reviewed in \citet{biver2019}. However, in this comet, methanol is very abundant, almost a factor of 20 more abundant than ethanol, which is probably the main reason for the deviating ratio.\\
Interestingly, diols tend to be more abundant than the respective mono-alcohol. This becomes obvious from the reports of \citet{biver2019} for ethanol vs. ethylene glycol (IUPAC standard name: 1,2-ethanediol) in both JFCs and OCCs, cf. also Fig. \ref{fig:comp-com}. Comet C/2014 Q2 (Lovejoy), where ethanol is almost twice as abundant as ethylene glycol, seems to be a bit of an exception. From our data, we find an order of magnitude more of the diol, too. The same applies to the next higher homologous series members, i.e., propanediol is more abundant than \textit{n}-propanol by almost a factor of 5. If the branched and unbranched isomer of the mono-alcohol are summed up, the same factor is only 3. Methanediol (CH$_4$O$_2$) is not observed in our data, i.e., there is no signal at \textit{m/z} = 48 where M would be expected. However, NIST does not list fragmentation data for this molecule, which might point towards increased instability. Indeed, it is known that geminal diols are prone to decay into water and the respective ketone or aldehyde, but the formation of methanediol from energetically processed methanol-oxygen ices and its gas phase stability could recently be demonstrated \citep{zhu2022}. The larger diols – namely, ethanediol and propanediol – are observed in the sterically more favorable vicinal 1,2-form. The isolated 1,3-propanediol loses water and produces strong signals on \textit{m/z} = 58 and 57 but no relevant M signal. However, the corresponding measured intensity is already well explained by the presence of acetone and propanal and hence possible contributions of the isolated diol are marginal. The largest diol that is very likely present in comet 67P's coma is the benzene-derivative hydroquinone and those structural isomers where the hydroxy functions are located at different positions on the benzene ring.

Otherwise it’s fine.

\subsection{Ethers (R--O--R)}\label{subsec:eth}
Due to their importance to biomolecules, \citet{barnum2022} recently performed an extensive search of cyclic ethers and other heterocycles toward the Taurus Molecular Cloud (TMC-1). They used the GOTHAM (GBT Observations of TMC-1: Hunting Aromatic Molecules) collaboration survey and reported upper limits for three O-bearing heterocycles, namely, ethylene oxide (C$_2$H$_4$O), furan (C$_4$H$_4$O), and benzofuran (C$_8$H$_6$O). Ethylene oxide, also known as oxirane, has been first identified in Sgr B2 by \citet{dickens1997}. Since then, it has been observed in several other interstellar environments. Also, its methylated form, propylene oxide also known as 2-methyloxirane (\textit{c}-CH$_3$C$_2$H$_2$O), was detected towards Sgr B2 \citep{mcguire2016}. As summarized in \citep{barnum2022}, furan has been searched for in several sources previously, but only upper limits were reported to date. The most constraining one was derived from the GOTHAM line survey by \citet{barnum2021} and corresponds to 1$x$10$^{12}$ cm$^{-2}$. \citet{barnum2022} interpret their null detections of the searched heterocycles as to point towards a unique chemistry of these species compared to the pure carbocycles and hypothesize that their absence in TMC-1 may be a matter of reaction dynamics, given their high terrestrial stability.\\
A very different picture presents itself based on this work: For the first time, we can confirm the presence of significant amounts of extraterrestrial O-bearing heterocycles. Especially the 5-membered furan-derivatives are found to be abundant. Interestingly, dehydrogenated species seem to be less abundant than fully hydrogenated ones. 6-Membered pyran-based species as well as 6-membered rings with two O-atoms (dioxanes) are very likely present but much less common. Out of the 19 identified ethers, 12 are heterocyclic molecules and, accordingly, the relative sum abundance of the cyclic ethers outweighs that of the linear ones by almost a factor of 3, see Table \ref{tab:comp-met}. This imbalance is expected to be partially due to the methodological bias that chains tend to yield less stable M ions than rings and, hence, are more difficult to identify. However, our data does leave limited room for the presence of additional species, given the 30\% and 50\% error margins of the observed intensities, respectively. Additional heterocyclic molecules likely present in comet 67P that do not classify as ethers are two lactones (see above under Carboxylic acids and carboxylate esters) and furanmethanol (see above under Alcohols). Interestingly, not the simple dehydrogenated heterocycles like furan and pyran, but the functionalized and hydrogenated ones seem to be more abundant. Our findings from in situ comet observations hence confirm the suggestion of \citet{miksch2021} that the hydrogenated versions of heterocycles might be promising candidates for future interstellar searches. Based on theoretical work, these authors report that, under interstellar conditions, many heterocycles should be slowly reduced by abundant hydrogen atoms. In terms of O-bearing heterocycles, especially 2,3-dihydrofuran and 2,5-dihydrofuran should be abundant, likely more abundant than furan itself. This is consistent with our findings which, furthermore, imply that the fully hydrogenated species, THF, is even more abundant than 2,3-dihydrofuran and 2,5-dihydrofuran. A similar situation is found for the two lactone heterocycles (see above under \ref{subsec:acids}), where the fully hydrogenated $\gamma$-butyrolactone is more abundant than the dehydrogenated furanone. Also, tetrahydropyran is found to be abundant relative to methanol, but only about half as abundant as THF. Unfortunately, missing reference data do not allow for the clear identification of pyran itself and no direct comparison of the hydrogenated and the non-hydrogenated isomer is possible. Despite the fact that methylated or sometimes ethylated molecules seem to play a significant role, it must be stressed that the locations of the alkyl groups generally are not well constrained by mass spectrometry and the simultaneous presence of multiple isomers seems likely. This is true not only for the case of heterocycles but also for the other chemical groups of molecules discussed above. Moreover, the relative abundance of heterocyclic species is compared with their pure hydrocarbon counterparts. While \citet{barnum2022} find significant depletion of heterocycles relative to the pure carbocycles in TMC-1, we cannot confirm this depletion based on our data from comet 67P. In TMC-1 benzofuran is at least 10 times less abundant than indene. However, from our data analyzed for this work, both these species are similarly abundant in comet 67P. The same picture we find for furan and cyclopentadiene. The abundance of all four molecules is slightly above 1\% relative to methanol.\\
To the best of our knowledge, ethers have not been studied in detail in the Murchison meteorite or any other chondrite yet.

\subsection{Peroxides (R--OO--R)}\label{subsec:pero}
From our data, diethyl peroxide seems to be present in 67P's coma, while dimethyl peroxide may or may not. Also, the inorganic hydrogen peroxide was observed based on DFMS data collected when the comet was still far from its perihelion \citep{bieler2015}. Overall, the chemical class of peroxide does not seem to be abundant among complex organic molecules, which may be due to the low stability of the peroxide bond. Peroxides have not been identified to date in the ISM, in other comets, or in Murchison SOM as far as we know. Possibly, the low stability of the O--O bond impairs the detection in meteoritic SOM extracts.


\section{Summary and conclusions}\label{sec:sumconc}
In this work, we revisited 67P's inventory of O-bearing complex organic molecules previously studied by \citet{schuhmann2019a}, applying an Occam's razor-based approach to achieve deconvolution of the complex sum fragmentation pattern observed on 3 August 2015 by Rosetta's high-resolution mass spectrometer DFMS. Despite methodological limitations, such as the incompleteness of the NIST mass spectrometry data base used as reference in this work and ambiguities related to structural isomerism, we derive a minimum of 63 C$_n$H$_m$O$_x$ candidate molecules, roughly quintupling the number of species reported in \citet{schuhmann2019a}. However, additional contributions from low-abundance complex organics, evading mass-spectrometric detection, are likely. From a thorough comparison of our findings to the well-studied Murchison SOM \citep{botta2002,schmitt-kopplin2010,alexander2017} as well as to the ISM \citep{mcguire2022} and other comets \citep{biver2019}, we are able to conclude the following:

   \begin{itemize}
	    \item  Comet 67P's O-bearing complex organics expose diverse chemical functionalities from carboxylic acids and carboxylate esters via alcohols, aldehydes, ketones, to ethers. Deviations from the Murchison meteoritic inventory \citep{botta2002,schmitt-kopplin2010,alexander2017} can be understood when taking into account that methodological bias resulting from the usually acidic meteorite sample work-up processes \citep{martins2015} mainly affects the quantification of carboxylic acids, esters, and alcohols. In addition, parent body (aqueous) alteration is thought to lead to aggregation of smaller ('more pristine') molecules or monomers into larger ('less pristine') (macro)molecular structures (e.g., \citet{alexander2017}), possibly explaining a depletion of small species in meteorites as compared to comets. For instance, we cannot confirm any signal to indicate the presence of molecules bearing three or more oxygen atoms (e.g., dicabroxylic acids, hydroxy carboxylic acids, or sugars) in comet 67P. Those species, which seem abundant in meteoritic SOM \citep{alexander2017}, are hypothesized to form during aqueous alteration in parent bodies.
			\item Compared to other comets reviewed in \citet{biver2019}, C$_n$H$_m$O$_x$ species in comet 67P tend to be more abundant relative to methanol. However, this could be the effect of data selection (very dusty time period) and local enhancement due to dust sublimating near DFMS' ionization chamber. However, there are also parallels: Our findings that non-geminal diols are more abundant than the corresponding mono-alcohols, for instance, are consistent with ground-based observations of other comets where ethanediol is consistently more abundant than ethanol \citep{biver2019}. Despite the fact that production of methanediol from energetically processed methanol-oxygen ices as well as gas phase stability has been demonstrated \citep{zhu2022}, and just like other comet observations \citep{biver2019}, we cannot confirm its presence in 67P's coma. The reason may be this molecule's short lifetime and its proneness to undergo dehydration and form formaldehyde. 
	    \item We find significantly more hydrogenated than dehydrogenated heterocycles, which is consistent with theoretical work on heterocycles of \citet{miksch2021} and has been reported also for the case of carbocycles in 67P's coma by \citet{haenni2022}. Our findings may have implications for future radiofrequency or infrared spectroscopic observational campaigns from both ground-based (e.g., with the Green Bank Telescope) and space-based facilities (e.g., with the James Webb Space Telescope). For example, the millimeter wave spectrum of furan in the frequency range of 75--295 GHz was investigated by \citet{barnum2021} and used to derive an upper limit in the cold core TMC-1. If the ISM ratio of furan to its dehydrogenated counterpart THF is comparable to the one we derive in this work for comet 67P, a search for THF might be more promising. The rotational spectrum of THF was studied in \citet{meyer1999} and \citet{mamleev2001}, and the pseudorotational band $n = 0 \rightarrow n = 2$ was investigated in the 170--360 GHz range by \citet{engerholm1969} and \citet{melnik2003}. However, the rotational spectrum of THF is complex as may be the identification of this molecule in the ISM. In favor of the radio line searches, THF does have a considerably larger dipole moment than its dehydrogenated counterpart.
	    \item For the two cases where heterocycles (this work) are directly comparable to carbocycles \citep{haenni2022}, we find that both variants of molecules exhibit similar relative abundances. In contrast, \citet{barnum2022} have reported at least an order of magnitude difference between indene and its heterocyclic pendent benzofuran, the latter being the less abundant one. Overall, C$_n$H$_m$O$_x$ species are clearly less abundant than pure hydrocarbons also in comet 67P's coma \citep{haenni2022}.
	    \item From our work, cyclic species seem to be the more suitable candidates to explain the data than acyclic ones, which is in contrast to ISM studies where clearly prolate molecules are dominantly found \citep{mcguire2022,mccarthy2021}. Notably, while for the latter sometimes no fragmentation data can be found on NIST (e.g., polyynes) or the M signal tends to be weak to absent (e.g. carbon chain species), their functionalized -- commonly with a cyano group -- variants possess a pronounced permanent dipole moment and are thus easy to target in space via their rotational spectra. This was discussed in more detail in \citet{haenni2022}. As far as we know, for meteoritic SOM, the overall ratio of cyclic to acyclic species has not been studied in detail.
	    \item We could also study the abundance of branched versus unbranched alkyl chain species: In contrast to analysis of Murchison SOM \citep{jungclaus1976}, we clearly identify the primary alcohol \textit{n}-propanol, most likely together with its secondary counterpart isopropanol. The estimated abundance ratio of these two species is roughly 4, suggesting that secondary alcohol is the dominant one. However, a possible contribution of methoxyethane could lower this estimated factor. \citet{belloche2022} report a slight dominance of the normal over the iso-version while simulations of branched carbon chain chemistry in Sgr B2(N) seem to indicate that the degree of branching rises with molecular size \citep{garrod2017}. To understand how this ratio might be variable for different classes of molecules in comet 67P, further investigations both in the laboratory and in space are required. Having a full organic inventory as accessible by Rosetta's high-resolution mass spectrometer DFMS on hand for comet 67P, in the future, a more detailed investigation of, e.g., branched versus straight chain species will become possible allowing for a more detailed comparison to modeling work done for the ISM (cf. \citet{garrod2017}).
   \end{itemize}

It is still unclear whether the complex organics targeted in this study form from small precursors via bottom-up pathways or rather via top-down mechanisms from decomposing larger PAHs or macromolecular structures in carbonaceous dust grains, as discussed by \citet{burkhardt2021} or \citet{lee2019}. Laboratory work suggests that PAHs embedded in astrophysical ice analogs are not only hydrogenated but also oxygenated when subjected to Ly-$\alpha$ radiation \citep{gudipati2012}. According to \citet{bernstein1999}, UV irradiation of PAHs in simulated interstellar ices  also seems to provide probable formation pathways towards ethers as well as alcohols and quinones. However, our study of pristine cometary matter can only add significantly to that ongoing debate in combination with laboratory and modelling efforts tackling the formation of the complex organics in the ISM, including the ratios of hydrogenated vs. dehydrogenated, straight vs. branched, and cyclic vs. acyclic species. While we are planning to further explore comet 67P's chemical complexity by extending our work to species containing heteroelements other than oxygen, we strongly encourage such future efforts.


\begin{acknowledgements}
We gratefully acknowledge the work of the many engineers, technicians and scientists involved in the Rosetta mission and in the planning, construction, and operation of the ROSINA instrument in particular. Without their contributions, ROSINA would not have produced such outstanding data and scientific results. Rosetta is an ESA mission with contributions from its member states and NASA. Work at the University of Bern was funded by the Canton of Bern and the Swiss National Science Foundation (200020 182418). S.F.W. acknowledges financial support of the SNSF Eccellenza Professorial Fellowship PCEFP2\_181150. M.R.C. acknowledges support from NASA grants 80NSSC18K1280 and 80NSSC20K0651. J.D.K. acknowledges support from the Belgian Science Policy Office. Especially, we thank Prof. Dr. B. McGuire for discussing our work with us from an ISM perspective.
\end{acknowledgements}

\bibliographystyle{aa} 
\bibliography{2023_Haenni_CHOs} 

\begin{appendix} 

\onecolumn

\section{Overview of candidate molecules}\label{app:1}

\begin{longtable}{p{0.40\textwidth}p{0.12\textwidth}p{0.08\textwidth}p{0.17\textwidth}p{0.15\textwidth}}
  \caption{List of O-bearing neutral and non-radical organic molecules$^{(a)}$ identified as best candidates to explain the 3 August 2015 overall DFMS mass spectrum.}\\
 \toprule
Molecule (identifier)                                          & Sum formula          & Mass (Da)  & Level-of-confidence indicator    & Abundance rel. to methanol (\%) \\
\midrule
\multirow{1}{*}{Cometary O-bearing reference molecules}\\ \hline
Carbon monoxide (no. 1)                                        & CO                   & 28         & 3      & 292      \\
Carbon dioxide (no. 51)                                        & CO$_2$               & 44         & 3      & 1071     \\ \hline
\multirow{1}{*}{Carboxylic acids (R--COOH) and carboxylate esters (R--COO--R)}\\ \hline
Formic acid (no. 52)                                           & CH$_2$O$_2$          & 46         & 3      & 36       \\
Acetic acid (no. 54)                                           & C$_2$H$_4$O$_2$      & 60         & 3      & 11       \\
Methyl formate (no. 56)                                        & C$_2$H$_4$O$_2$      & 60         & 3      & 44       \\
2-Propenoic acid (no. 58)                                      & C$_3$H$_4$O$_2$      & 72         & 3      & 6.3      \\
Propanoic acid (no. 59)                                        & C$_3$H$_6$O$_2$      & 74         & 3      & 39       \\
2(3H)-Furanone (no. 63)                                        & C$_4$H$_4$O$_2$      & 84         & 2      & 12       \\
Cyclopropanecarboxylic acid (no. 64)                           & C$_4$H$_6$O$_2$      & 86         & 1      & 12       \\
$\gamma$-Butyrolactone (no. 65)                                & C$_4$H$_6$O$_2$      & 86         & 1      & 66       \\
Butanoic acid (no. 68)                                         & C$_4$H$_8$O$_2$      & 88         & 2      & 13       \\
Cyclopropanecarboxylic acid methyl ester (no. 73)              & C$_5$H$_8$O$_2$      & 100        & 2      & 2.5      \\
Propanoic acid ethyl ester (no. 74)                            & C$_5$H$_{10}$O$_2$   & 102        & 2      & 15       \\
Benzoic acid (no. 77)                                          & C$_7$H$_6$O$_2$      & 122        & 3      & 25       \\ \hline
\multirow{1}{*}{Aldehydes (R--CHO) and ketones (R--CO--R)}\\ \hline
Formaldehyde (no. 2)                                           & CH$_2$O              & 30         & 3      & 206      \\
Ketene (no. 4)                                                 & C$_2$H$_2$O          & 42         & 3      & 71       \\
Acetaldehyde (no. 5)                                           & C$_2$H$_4$O          & 44         & 3      & 150      \\
2-Propynal (no. 8)                                             & C$_3$H$_2$O          & 54         & 2      & 1.7      \\
2-Propenal (no. 9)                                             & C$_3$H$_4$O          & 56         & 2      & 29       \\
Propanal (no. 11)                                              & C$_3$H$_6$O          & 58         & 3      & 48       \\
Acetone (no. 10)                                               & C$_3$H$_6$O          & 58         & 3      & 118      \\
Glyoxal (no. 53)                                               & C$_2$H$_2$O$_2$      & 58         & 2      & 2.5      \\
Glycolaldehyde (no. 55)                                        & C$_2$H$_4$O$_2$      & 60         & 3      & 44       \\
Butanal (no. 16)                                               & C$_4$H$_8$O          & 72         & 3      & 12       \\
2,4-Cyclopentadiene-1-one (no. 19)                             & C$_5$H$_4$O          & 80         & 2      & ?$^{(b)}$   \\
3-Furaldehyde (no. 71)                                         & C$_5$H$_4$O$_2$      & 96         & 1      & 2.2      \\
Benzaldehyde (no. 28)                                          & C$_7$H$_6$O          & 106        & 3      & 4.1      \\
4-Methylbenzaldehyde (no. 33)                                  & C$_8$H$_8$O          & 120        & 2      & 4.3      \\
6-methyl-3,5-heptadien-2-one (no. 35)                          & C$_8$H$_{12}$O       & 124        & 1      & 8.3      \\
2,6-Dimethylcyclohexanone (no. 36)                             & C$_8$H$_{14}$O       & 126        & 1      & 19       \\ \hline
\multirow{1}{*}{Alcohols (R--OH)} \\ \hline
Methanol (no. 3)                                               & CH$_4$O              & 32         & 3      & 100$^{(c)}$ \\
Ethanol (no. 7)                                                & C$_2$H$_6$O          & 46         & 3      & 4.1      \\
Isopropanol (no. 12)                                           & C$_3$H$_8$O          & 60         & 2      & 19       \\
\textit{n}-Propanol (no. 13)                                   & C$_3$H$_8$O          & 60         & 3      & 5.0      \\
Ethylene glycol (no. 57)                                       & C$_2$H$_6$O$_2$      & 62         & 3      & 50       \\
2-Butanol (no. 18)                                             & C$_4$H$_{10}$O       & 74         & 2      & 8.3      \\
1,2-Propanediol (no. 62)                                       & C$_3$H$_8$O$_2$      & 76         & 2      & 24       \\
2-Methoxyethanol (no. 60)                                      & C$_3$H$_8$O$_2$      & 76         & 3      & 27       \\
Phenol (no. 25)                                                & C$_6$H$_6$O          & 94         & 2      & 7.7      \\
2-Furanmethanol (no. 72)                                       & C$_5$H$_6$O$_2$      & 98         & 3      & 18       \\
Benzyl alcohol (no. 29)                                        & C$_7$H$_8$O          & 108        & 1      & 4.1      \\
Hydroquinone (no. 76)                                          & C$_6$H$_6$O$_2$      & 110        & 2      & 2.6      \\
2-Methylcyclohexanol (no. 31)                                  & C$_7$H$_{14}$O       & 114        & 1      & 30       \\ \hline
\multirow{1}{*}{Ethers (R--O--R)} \\ \hline
Dimethyl ether (no. 6)                                         & C$_2$H$_6$O          & 46         & 3      & 3.1      \\
Furan (no. 14)                                                 & C$_4$H$_4$O          & 68         & 2      & 1.7      \\
2,3-Dihydrofuran (no. 15)                                      & C$_4$H$_6$O          & 70         & 2      & 8.3      \\
Tetrahydrofuran (no. 17)                                       & C$_4$H$_8$O          & 72         & 3      & 64       \\
Methylal (no. 61)                                              & C$_3$H$_8$O$_2$      & 76         & 2      & 4.0      \\
2-Methylfuran (no. 20)                                         & C$_5$H$_6$O          & 82         & 3      & 11       \\
2,3-Dihydro-4-methylfuran (no. 21)                             & C$_5$H$_8$O          & 84         & 3      & 20       \\
Tetrahydropyran (no. 22)                                       & C$_5$H$_{10}$O       & 86         & 3      & 25       \\
1-Ethoxypropane (no. 23)                                       & C$_5$H$_{12}$O       & 88         & 3      & 3.9      \\
2-Methoxy-2-methylpropane (no. 24)                             & C$_5$H$_{12}$O       & 88         & 3      & 0.2      \\
1,3-Dioxane (no. 66)                                           & C$_4$H$_8$O$_2$      & 88         & 3      & 6.2      \\
1,4-Dioxane (no. 67)                                           & C$_4$H$_8$O$_2$      & 88         & 3      & 13       \\
1-Ethoxy-1-methoxyethane (no. 69)                              & C$_4$H$_{10}$O$_2$   & 90         & 2      & 19       \\
2,5-Dimethylfuran (no. 26)                                     & C$_6$H$_8$O          & 96         & 3      & 11       \\
3-Methoxycyclopentene (no. 27)                                 & C$_6$H$_{10}$O       & 98         & 3      & 33       \\
4-Methyl-1,3-dioxane (no. 75)                                  & C$_5$H$_{10}$O$_2$   & 102        & 2      & 20       \\
2,3,5-Trimethylfuran (no. 30)                                  & C$_7$H$_{10}$O       & 110        & 2      & 9.3      \\
Benzofuran (no. 32)                                            & C$_8$H$_6$O          & 118        & 3      & 1.9      \\
Ethoxybenzene (no. 34)                                         & C$_8$H$_{10}$O       & 122        & 1      & 19       \\ \hline
\multirow{1}{*}{Peroxides (R--OO--R)} \\ \hline
Diethyl peroxide (no. 70)                                      & C$_4$H$_{10}$O$_2$   & 90         & 2      & 3.8      \\
\bottomrule
\multirow{1}{0.9\textwidth}{\tablefoot{$^{(a)}$ Molecules are subdivided into groups according to their chemical functionalities like in main text of Section \ref{sec:disc}, starting always with the molecule with the smallest molecular weight. CO and CO$_2$ are excluded because they are not organic and do not belong in any of the groups. $^{(b)}$ No NIST mass spectrum available for any of the structural isomers of with the sum formula C$_5$H$_4$O. The proposed molecule seems a plausible candidate but no ARM can be estimated due to the missing fragmentation information. $^{(c)}$ Definition. Abundance estimates are normalized relative to methanol.}}
\label{tab:overview}
  \end{longtable}




\end{appendix}

\end{document}